\documentclass[a4paper]{aa}
\usepackage{xspace}
\usepackage{graphicx}
\usepackage{color}
\usepackage{txfonts}
\usepackage{natbib}
\usepackage{lscape}
\usepackage{longtable}
\usepackage{balance}
\usepackage{placeins}
\usepackage{paralist}
\usepackage{xtab,booktabs}
\usepackage{multirow}
\usepackage{bbold}
\usepackage{ulem}

\usepackage[breaklinks=true,bookmarks=true,hypertexnames=false,bookmarksnumbered=true,bookmarksopen=false,bookmarks=true,bookmarksnumbered=true]{hyperref}
\usepackage[]{color}

\definecolor{Red}{rgb}{0.65,0.08,0.05}

\definecolor{Purple}{rgb}{0.45,0.05,0.65}

\definecolor{Blue}{rgb}{0.15,0.15,0.75}

\usepackage{array,multirow,makecell}
\setcellgapes{1pt}
\makegapedcells
\newcolumntype{R}[1]{>{\raggedleft\arraybackslash }b{#1}}
\newcolumntype{L}[1]{>{\raggedright\arraybackslash }b{#1}}
\newcolumntype{C}[1]{>{\centering\arraybackslash }b{#1}}

\newcommand{\msun}{{\rm M}_{\odot}}

\newcommand{\h}{\,h}
\newcommand{\hmm}{\h^{-1}}
\newcommand{\hmmsun}{\hmm\msun}
\newcommand{\kpc}{\, {\rm kpc}}

\newcommand{\Mpc}{\, {\rm Mpc}}
\newcommand{\hmMpc}{\hmm\Mpc}
\newcommand{\der}{{\rm d}}

\newcommand*\Email[1]{E-mail: {\aa@emailfont #1}}

\newcommand{\hagn}{Horizon-AGN}
\newcommand{\ramses}{{\tt RAMSES}}

\newcommand\footnoteref[1]{\protected@xdef\@thefnmark{\ref{#1}}\@footnotemark}

\begin{document}

\title{Weak lensing in the \hagn~simulation lightcone }{
\subtitle{Small scale baryonic effects}

\author{
  C.~Gouin\inst{1,2}\thanks{E-mail:~\tt{celine.gouin@ias.u-psud.fr}}, R.~Gavazzi\inst{1}, C.~Pichon\inst{1,3,4},
  Y.~Dubois\inst{1},  C.~Laigle\inst{5}, N.~E.~Chisari\inst{5}, S.~Codis\inst{1}, J.~Devriendt\inst{5}, S.~Peirani\inst{6}
}

\institute{
Institut d'Astrophysique de Paris, UMR7095 CNRS \& Sorbonne Universit\'e, 98bis Bd Arago, F-75014, Paris, France
\label{inst1}
 \and
Institut d'Astrophysique Spatiale, CNRS/Universit\'e Paris-Sud, Universit\'e Paris-Saclay, B\^atiment 121, 91405 Orsay, France
\label{inst2}
\and
Korea Institute of Advanced Studies (KIAS) 85 Hoegiro, Dongdaemun-gu, Seoul, 02455, Republic of Korea
\label{inst3}
\and
 Institute for Astronomy, University of Edinburgh, Royal Observatory, Blackford Hill, Edinburgh, EH9 3HJ, United Kingdom
 \label{inst4}
\and
Sub-department of Astrophysics, University of Oxford, Keble Road, Oxford OX1 3RH, United Kingdom
\label{inst5}
 \and
Universit\'e C\^ote d'Azur, Observatoire de la C\^ote d'Azur, CNRS, Laboratoire Lagrange, Bd de l'observatoire, 06304 Nice, France
 \label{inst6}
}

\date{\today}


\abstract
{Accurate model predictions including the physics of baryons are required to make the most of the upcoming large cosmological surveys devoted to gravitational lensing. The advent of hydrodynamical cosmological simulations enables such predictions on sufficiently sizeable volumes.
}
{ 
Lensing quantities (deflection, shear, convergence)  and their statistics (convergence power spectrum, shear correlation functions, galaxy-galaxy lensing) are computed in the past lightcone built in the \hagn~hydrodynamical cosmological simulation, which implements our best knowledge on baryonic physics at the galaxy scale in order  to mimic galaxy populations over cosmic time.  
}
{ 
Lensing quantities are generated over a one square degree field of view by performing multiple-lens plane ray-tracing through the lightcone, taking full advantage of the 1 kpc resolution and splitting the line of sight over 500 planes all the way to redshift $z\sim7$. Two methods are explored (standard projection of particles with adaptive smoothing, and integration of the acceleration field) to assert a good implementation. The  focus is on small scales where baryons matter most.
}
{
Standard cosmic shear statistics are impacted at the 10\% level by the baryonic component for angular scales below a few arcmin. The galaxy-galaxy lensing signal, or galaxy-shear correlation function, is consistent with measurements for the redshift $z\sim0.5$ massive galaxy population. At higher redshift $z\gtrsim 1$, the impact of magnification bias on this correlation is relevant for separations greater than 1 Mpc.
}
{
This work is pivotal for all current and upcoming weak lensing surveys and represents a first step towards 
building a full end-to-end generation of lensed mock images from large cosmological hydrodynamical simulations.
}

   
\authorrunning{Gouin et al.}
\titlerunning{Raytracing through the \hagn~lightcone}

\maketitle

\section{Introduction}
Gravitational lensing has become a versatile tool to probe the cosmological model and scenarios of galaxy evolution. From the coherent distorsions, generated by the intervening matter along the line of sight, of the last scattering surface \citep[eg][]{Planck18-8} or intermediate redshift galaxies \citep{BS01,KilbingerReviewCS}, to the inner parts of massive galaxies \citep{Tre10}, lensing directly measures the fractional energy density in matter of the Universe. Since it does not rely on assumptions about the relative distribution between the galaxies and the underlying Dark Matter (DM), which drives the dynamical evolution of cosmological structures, weak lensing plays a key role in recent, ongoing or upcoming ground-based imaging surveys \citep[CFHTLenS, DES, KiDS, HSC, LSST,~][]{CFHTLens,DES,DES_CS,KIDS,HSC,LSST}. It is also at the center of the planned Euclid and WFIRST satellites \citep{Euclid,wfirst}. 

The statistical power of these experiments dramatically increases and drives on its way enormous efforts for the control of systematic effects. One of them concerns the accuracy to which theoretical predictions on the statistical properties of the matter distribution when it has evolved into the non-linear regime can be made on small scale. Arguably, cosmological N-body numerical simulations have been playing a key role in solving the complex dynamical evolution of DM on scales smaller than a few Mpc \citep[eg][]{SFW06}. The upcoming Euclid or LSST missions require an extreme accuracy on the matter density power spectrum and the associated covariances which may enter a likelihood analysis of these data. The effort is currently culminating with the Flagship simulation, for instance \citep{flagship17}. But it also motivated earlier very large simulations like Horizon-4$\pi$ \citep{Teyssier++09,Pichon++10}, DEUS \citep{DEUS10} or MICE \citep{MICE1}.
It has early been envisioned to propagate light rays through such dark matter simulations in order to reproduce the deflection and distorsions of light bundles in a lumpy universe.  The motivation is to derive lensing observables like convergence maps and 1-point PDFs of this field or its topological properties (peaks, voids...) or 2-point shear correlation functions \citep[eg][]{JSW00,Pichon++10,Ham01,Vale+03,H+S05,Hil++07,hil++09,Sato++09}. Since, much progress has been made on large and mildly non linear scales with the production of full sky maps with a few arc minutes angular resolution \citep[eg][]{MICE3,Giocoli+16,Takahashi++17}.

In order to make the most of the upcoming surveys, the matter distribution for Fourier modes as large as $k\sim 10 h \Mpc^{-1}$ must be predicted to the percent accuracy, which nowadays still represents a challenge \citep{Schneider++16}. Furthermore, at those scales, the physics of baryons can differ from the dynamics of DM and, even though, it amounts for $\sim 17\%$ of the total cosmological matter budget, it has to be taken into account \citep[OWLS simulation][]{VanDaalen2011}.
For weak lensing statistics, \citet{Semboloni++11} showed that the modelling of the 2-point shear correlation function can be significantly biased, should the baryons be simply treated like the collision-less DM. Even the number of convergence peaks itself is altered by baryons but to a 
 lesser extent than the power-spectrum  \citep{Yang++13}.

Recently, significant progress has been made on hydrodynamical simulations which are now able to reproduce a morphological mix of galaxies in a cosmological context, by considering baryonic physics such as radiative cooling, star formation, and feedback from supernovae and Active Galactic Nuclei (AGN). Despite the tension between the high resolution needs to properly describe the galaxies formed at the center of DM halos and the necessity to simulate sizeable cosmological volumes, recent simulations, such as \hagn~\citep{Dubois++14}, Illustris/Illustris-TNG \citep{illustris,Pillepich18}, or EAGLE \citep{Eagle}, have now reached volumes of order $100 \Mpc$ on a side
 and resolution of order $1 \kpc$. This opens the possibility to quantifying the effect of baryons (experiencing adiabatic pressure support, dissipative cooling, star formation, feedback...) on the total matter distribution and its impact on lensing cosmological observables \citep[see e.g.][]{VanDaalen2011, tenneti+15, hellwing+16,IllustrisTNG,Chisari++18}. Prescriptions to account for this effect \citep[eg][]{Semboloni++13,S+T15,Mead++15,R+T17} have been explored and some start to be incorporated in cosmic shear studies \citep[KIDS:][]{Hildebrandt++17}.

In this paper, we further investigate the impact of baryons on lensing observables in the \hagn~simulation. By taking advantage of the lightcone generated during the simulation run, we are able to fully account for projection effects (mixing physical scales) and small scale non-linearities occurring in the propagation of light rays (eg, Born approximation, lens-lens coupling, shear -- reduced shear corrections) which may be boosted by the steepening of the gravitational potential wells due to cooled gas sinking at the bottom of DM halos. Hence, this extends the analysis of \citet{Chisari++18} who mostly focused on the effect of baryons on the three-dimensional matter power spectrum, compared the \hagn~results with those of Illustris, OWLS, EAGLE and Illustris-TNG, found a broad qualitative agreement. 
The common picture is that hot baryons which are prevented from sinking into halos like DM, induce a deficit of power inside halos (in a proportion of order $\Omega_{\rm b}/\Omega_{\rm M}$) and, at yet smaller scales ($k \gtrsim 30 h \Mpc^{-1}$), baryons in the form of stars (and to a lesser extent cooled gas) dramatically boost the amplitude of density fluctuations.
However, even though those results seem to converge from one simulation to another, they substantially depend on the assumptions about sub-grid physics, and in particular about AGN feedback. 

Beside those encouraging successes at quantifying the nuisance of baryons on cosmological studies, hydrodynamical simulations entail a wealth of information on the relation between galaxies or galaxy properties and the halo they live in. It is, thereby, a way to understand the large scale biasing of these galaxies with respect to the overall total matter density field.
We also explore the small scale relation between galaxies and their surrounding gravitational potential sourcing the lensing deflection field. In particular, the correlation between galaxies and the tangential distortion of background sources (so-called Galaxy-Galaxy Lensing signal, GGL) has proven being a way to constrain the galaxy-mass correlation function \citep[eg][]{brainerd96,guzik01,mandelbaum06,2013MNRAS.432.1544M,Lea++12,2014MNRAS.437.2111V,2015MNRAS.447..298H,2015MNRAS.449.1352C}. In this vein, \citet{Velliscig++17} recently showed that the GGL  around $z\sim 0.18$ galaxies in the EAGLE simulation is consistent with the GAMA+KiDS data \citep{Dvornik++17}. 

Finally,  subtle observational effects entering GGL by high redshift deflectors ($z \gtrsim 0.8)$ are investigated from the lensing information over the full past lightcone of the \hagn~simulation. The magnification bias affecting the selection of deflectors \citep{Z+H08} complicates the interpretation of GGL  substantially. Currently, no such high-z lens sample has been studied because of the scarcity of even higher faint lensed sources carrying the shear signal but the situation may change with Euclid. Its slit-less grism spectroscopy will provide a large sample of H$\alpha$ emitters in the $0.9\le z\le 1.8$ redshift range. A thorough understanding of the clustering properties of this sample may be achieved with the GGL measurement of this sample by using the high-z tail of the shape catalogue obtained with the VIS imager. Some raytracing through cosmological simulations \citep{hil++09,MICE3} had briefly mentioned some aspects of the problem of magnification bias raised by \citet{Z+H08}. The \hagn~lightcone is a good opportunity to  quantify those effects in order to correctly interpret upcoming GGLs. 
In this paper, cosmic shear or GGL quantities are directly measured from the lensing quantities obtained by  ray-tracing methods. They are not inferred from the shape of galaxies as is done in observations.
A forthcoming paper will present  the generation of mock wide-field images including lensing distortions from the full view of \hagn~lightcone and the light emission predicted for the simulated stars, taking us one step closer to a full end-to-end generation of mock lensing observations.

The paper is organised as follows.
Sect.~\ref{sec:simintro} presents the \hagn~hydrodynamical simulation, the structure of its lightcone and some properties of the galaxy population, therein.
Sect.~\ref{sec:method} describes the   implemented methods to generate the deflection field on thin lens planes and to propagate light rays through them.
 Sect.~\ref{sec:cosmicshear},  describes the 1-point and 2-point statistics of the resulting convergence and (reduced-)shear fields. The validity of the raytracing method is quantified by comparing our results with independent methods.
Sect.~\ref{sec:GGL} measures the GGL around the galaxies in the \hagn~simulation. A comparison with observations is made for low redshift deflectors. The problem of magnification bias  is investigated for future observations of high-z GGL.    Sect.~\ref{sec:conclusions} wraps up.

\section{The Horizon-AGN simulation lightcone}\label{sec:simintro}

\subsection{Characteristics}\label{ssec:sim0}

The \hagn~simulation is a cosmological hydrodynamical simulation performed with \ramses~\citep{teyssier02}. The details of the simulations can be found in \cite{Dubois++14}. Let us first briefly summarise the main characteristics. \hagn~contains $1024^3$ dark matter particles with a mass resolution of $8 \times 10^7\hmmsun$, in a box of comoving size $L_{\rm box} = 100\,\hmMpc$ on a side. The gravity and hydrodynamics are treated in \ramses~with a {\it multiscale} approach with adaptive mesh refinement (AMR): starting from a uniform $1024^3$ grid, cells are then adaptively refined when the mass inside the cell exceeds $8$ times the initial mass resolution. Cells are recursively refined (or de-refined according to the refinement criterion) down to a minimum cell size of almost constant 1 proper kpc (an additional level is triggered at each expansion factor $a=0.1,0.2,0.4,0.8$). The underlying cosmology is a standard $\Lambda$CDM model consistent with the WMAP7 data \citep{komatsuetal11}, with total matter density $\Omega_{\rm m}=0.272$, dark energy density $\Omega_{\rm \Lambda}=0.728$, amplitude of the matter power spectrum $\sigma_8=0.81$, baryon density $\Omega_{\rm b}=0.045$, Hubble constant of $H_0=70.4\, \rm km\, s^{-1}\, Mpc^{-1}$, and scalar spectral index $n_{\rm s}=0.967$. 

The evolution of the gas is solved on the \ramses~grid using a Godunov method with the approximate HLLC Riemann solver on the interpolated conservative hydrodynamical quantities, that are linearly interpolated at cell boundaries from their cell-centered values using a MinMod total variation diminishing scheme.
In addition, accurate models of unresolved sub-grid physics have been implemented.
The gas heating comes from a uniform UV background which started at the re-ionisation $z_{\rm reion} = 10$ \citep{haardt&madau96}. 
The cooling function of the gas follows \citet{Sutherland1993}, from H and He collision and from the contribution of other metals. 
Star formation is modelled following the Schmidt law \citep{Kennicutt1998}, with a constant star formation efficiency of $2\%$ per free fall time. It occurs when the density of the gas exceeds the threshold $0.1 \ \rm{H\, cm^{-3}}$.
The temperature at gas densities larger than $0.1 \ \rm{H\, cm^{-3}}$ is modified by a polytropic equation of state with polytropic index of $4/3$ and scaling temperature of $10^4\, \rm K$ \citep{Springel2003}. 
Stellar evolution is performed assuming a \citet{Salpeter1955} initial stellar mass function. The sub-grid physics also includes stellar winds and supernova feedback in the form of heating, metal enrichment of the gas, and kinetic energy transfer to the ambient gas \citep[see][ for more details]{Kaviraj++17}.
Finally, black holes (BH) are created when the gas density exceeds $0.1 \ \rm{H\, cm^{-3}}$, and when there is not other BH in the close environment.
They grow by direct accretion of gas following an Eddington-limited Bondi-Hoyle-Littleton accretion rate, and merger when BH binaries are sufficiently close.
The AGN feedback is treated by either an isotropic injection of thermal energy, or by a jet as a bipolar outflow, depending of the ratio between the Bondi and the Eddington accretion rates
\citep[see][for details]{duboisetal12agnmodel, Volonteri++16}. 

The past lightcone of the simulation was created on-the-fly as the simulation was running. Its geometry is sketched in Fig.~\ref{fig:cone-shape}. The opening angle of the cone is $2.25$ deg out to redshift $z=1$ and $1$ deg all the way to $z=8$. These two values correspond to the angular size of the full simulation box at these redshifts. We can therefore safely work in the flat sky (or infinitely remote observer) approximation.
Up to $z=1$, the volume of the cone is filled with $\sim7$ replicates of the box.  Between $z=0$ and $z=4$, the narrow cone contains $\sim14$ replicates of the box and the union of the two cones contains about $19$ copies. This should be kept in mind when quantifying the statistical robustness of our results. 

 In order to limit projection effects, a non-canonical direction is chosen for the past lightcone but, in order to preserve periodic boundary conditions between replicates, no random rotation is applied. Projection effects will still be present and induce characteristic spectral distortions on large scales which must be taken into account.
Particles and AMR cells were extracted on-the-fly at each coarse simulation time step (when all levels are synchronized in time as a factor 2 of subcycling is used between levels) of the simulation according to their proper distance to a fiducial observer located at the origin of the simulation box. The lightcone of the simulation thus consists in 22,000 portions of concentric shells. Each of them contains stellar, black hole, DM particles (with their position and velocity, mass and age) along with AMR Eulerian cells storing the gas properties (position, density, velocity, temperature, chemical composition, and cell size) and the total gravitational acceleration vector.

\begin{figure}[h]
\includegraphics[width=0.499\textwidth]{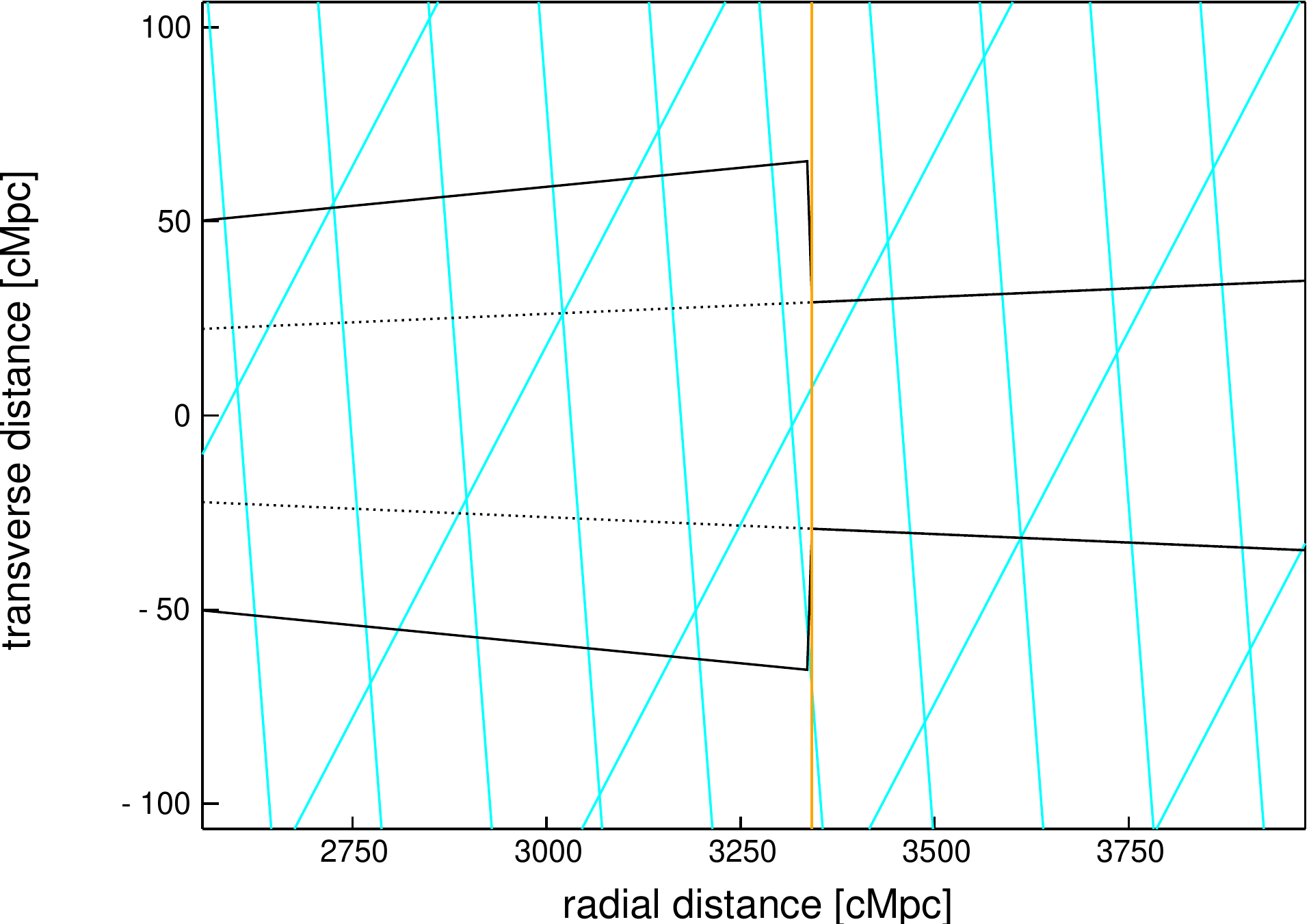}  
\caption{2D sketch of the past lightcone around redshift $z=1$ (orange vertical line). Each mesh is a replicate of the \hagn~simulation box (bounded with cyan lines). The tiling is performed all the way up to redshift $z\sim8$. }
\label{fig:cone-shape}
\end{figure}

\subsection{Properties of galaxies and host halos}\label{ssec:catalogs0}

The \textsc{ AdaptaHOP} halo finder \citep{aubert04} is run on the lightcone  to identify galaxies from the stellar particles distribution. Local stellar particle density is computed from the 20 nearest neighbours, and structures are selected with a density threshold  equal to 178 times the average matter density at that redshift. Galaxies resulting in less than 50 particles ($\simeq 10^8 \,{\rm M}_\odot$) are not included in the catalogue.
Since the identification technique is redshift dependent, \textsc{AdaptaHOP} is run iteratively on thin lightcone slices. Slices are overlapping to avoid edge effects (i.e. cutting galaxies in the extraction) and duplicate are removed. 
In a second step dark matter haloes have been extracted independently  from the dark matter particle distribution, with a density threshold of 80 times the average matter density, and keeping only haloes with more than 100 particles.  The centre of the halo is temporarily defined as the densest particle in the halo, where the density is computed from the 20 nearest neighbours. In a subsequent step, a sphere of the size of the virial radius is drawn around it and implement a shrinking sphere method (Power et al. 2003) to recursively find the centre of mass of the halo. In each iteration, the radius of the halo is reduced by 10 $\%$. The search is stopped when a sphere 3 times larger than our spatial resolution is reached. Each galaxy is matched with its closest halo. 

The simulation contains about $116,000$ galaxies and halos in the simulation box at $z=0$, with a limit of order $M_* \gtrsim 2\times10^9 \msun$.
These yields have been extensively studied in previous papers of the \hagn~series. For instance, \citet{Kaviraj++17} compared the statistical properties of the produced galaxies, showing a reasonable agreement with observed stellar mass functions all the way to $z\sim6$. The colour and star formation histories are also well recovered and so are the black hole -- bulge relations and duty-cycles of AGNs \citep{Volonteri++16}.

Following up on an earlier work \citep{Dub++13} focusing on a handful of zoomed galaxy simulations with \ramses, \citet{HorizonAGN} confirmed with a much greater statistical significance in \hagn, that the morphological diversity of galaxies is well reproduced (fraction of rotation- versus dispersion-supported objects, and how this dichotomy maps into the star forming versus quiescent dichotomy). Taking advantage of a parallel simulation run with the same initial conditions and in which the AGN feedback is turned off (Horizon-noAGN), the key role of the latter in shaping the galaxy morphology was emphasised. Furthermore, \citet{Peirani++17} studied the effect of AGN feedback on the innermost density profiles (stars, gas, DM, total) and found a good agreement of the density profile, size-mass relation and dark matter fraction inside the effective radius of galaxies with observations. In particular, \citet{Peirani++18} showed that the innermost parts of \hagn~galaxies are consistent with strong lensing observations of \citet{Sonnenfeld++13} and \citet{New++13,New++15}.

 Populating the lightcone yields  a volume limited sample of $1.73\times 10^6$ galaxies in the narrow 1 deg cone. However, a large fraction of the low mass high redshift galaxies would not be of much practical use in a flux limited survey as shown in Fig.~\ref{fig:gals} which plots the redshift dependent limit in stellar mass attained with several $i$-band apparent limiting magnitudes. This was obtained using the COSMOS2015 photometric catalogue of \citet{Laigle16}.

\begin{figure}[h]
\includegraphics[width=0.499\textwidth]{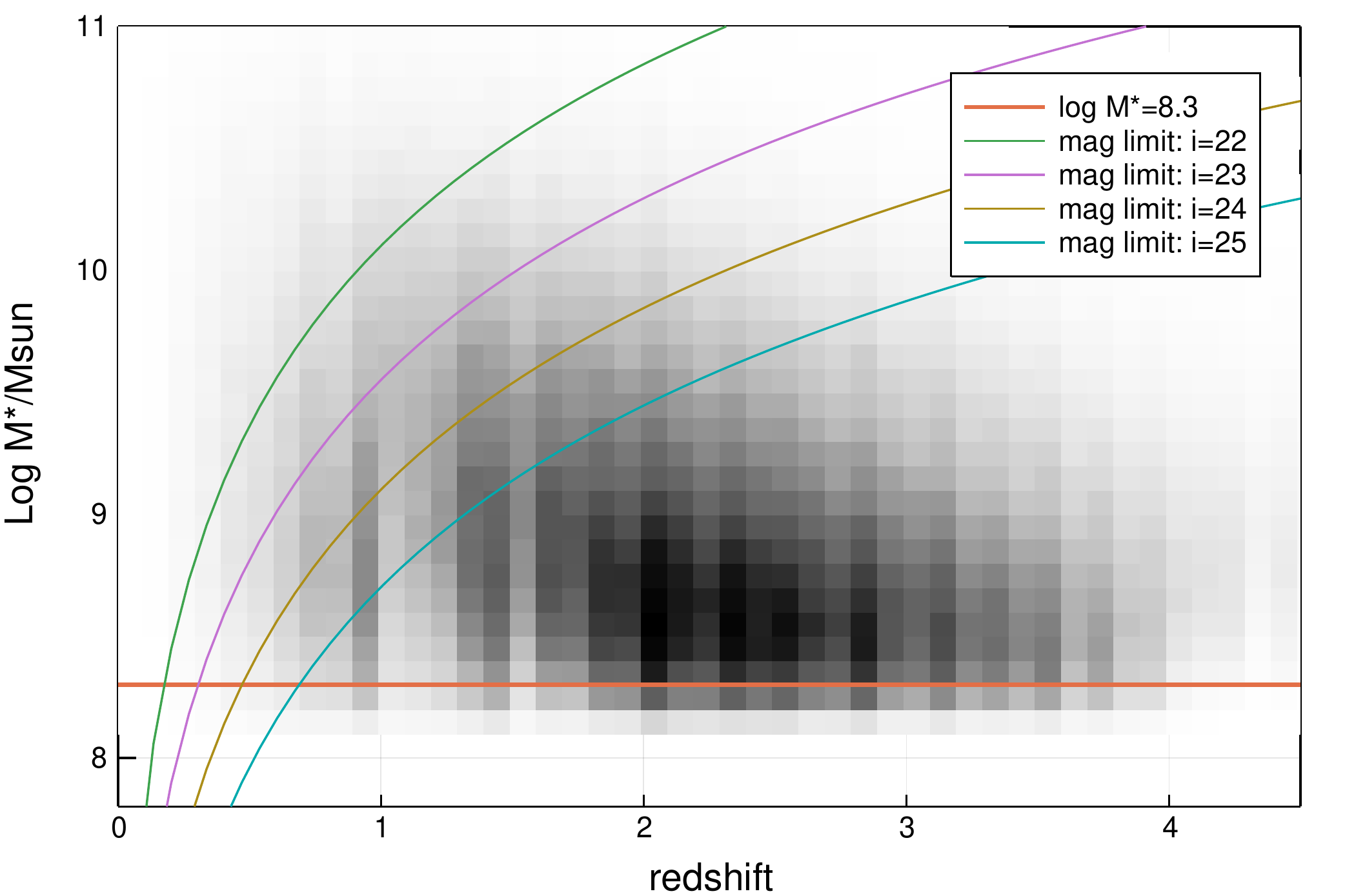}  
\caption{Distribution in the redshift -- stellar mass plane of the $1.7$ million galaxies in the \hagn~lightcone. For guidance the stellar mass limit for completeness is shown as well as fiducial cuts in mass one would obtain with a flux limited survey of various $i$ band limiting magnitudes.}
\label{fig:gals}
\end{figure}

\section{Raytracing through the lightcone}\label{sec:method}

After briefly describing the basics of the propagation of light rays in a clumpy universe and the numerical transcription of this formalism, 
let us now describe  the  the ray-tracing computation in the \hagn~lightcone. Our implementation of the multiple lens plane (but also the Born approximation) builds on similar past efforts \citep{Hil++08,GLAMER1,GLAMER2,Barreira++16}. It has been tailored for the post-treatment of the \hagn~past lightcone, but, provided the flat sky approximation holds, our implementation could readily be applied to any other \ramses~lightcone output \citep{Teyssier++09}.

As detailed below, two methods are investigated to infer deflection angles from either the distribution of various particle-like matter components or the total gravitational acceleration stored by \ramses. 
 The light rays are then propagated plane by plane (both within and beyond the Born approximation), for these two different estimates of the deflection field.

\subsection{The thin lens plane}
Let us define $\vec{\beta}$ the (un-perturbed and unobservable) source plane angular position and $\vec{\theta}$  the observed angular position of a light ray. Considering a unique, thin, lens plane, the relation between the angular position of the source $\vec{\beta}$, the deflection angle $\vec{\alpha}$ and the image $\vec{\theta}$ is simply:
\begin{equation}\label{eq:lenseq}
\vec{\beta} = \vec{\theta} -\frac{D_{\rm ls}}{D_{\rm s}} \vec{\alpha}(\vec{\theta})\,,
\end{equation}
where $D_{\rm ls}$ and $D_{\rm s}$, are the angular diameter distance between the source and the lens, and between the observer and the source, respectively.
The deflection angle $\vec{\alpha}(\vec{\theta})$ is obtained by integrating the gravitational potential $\Phi(\vec{r})$ along the line of sight (here, radial proper coordinate $x_3$)
\begin{equation}\label{eq:alphadef}
 \vec{\alpha} (\vec{\theta}) = \frac{2}{c^2} \int \vec{\nabla_{\perp}} \Phi(\vec{\theta},x_3) \ {\rm d} x_3\,.
\end{equation}
Hence, across a thin lens plane, the lensing potential $\phi(\vec{\theta})$ is related to the deflection field by the Poisson equation:
\begin{equation}\label{eq:poisson}
\Delta \phi = \vec{\nabla} . \vec{\alpha} \equiv 2 \kappa \,,
\end{equation}
where the convergence $\kappa$ is the projected surface mass density $\Sigma(\vec{\theta})$ in the lens plane expressed in units of the critical density $\Sigma_{\rm crit}$ 
\begin{equation}
\Sigma_{\rm crit}\, \kappa(\vec{\theta}) = \Sigma(\vec{\theta}) \equiv \int \rho(\vec{\theta},z)\, \der z\,.
\end{equation}
The critical density reads:
\begin{equation}\label{eq:scrit}
\Sigma_{\rm crit} = \frac{c^2}{4 \pi G} \frac{D_{\rm s}}{D_{\rm l}D_{\rm ls}}\,,
\end{equation}
with $D_{\rm l}$, the angular diameter distance between the observer and the lens. In the above equations, all distances and transverse gradients are expressed in physical (proper) coordinates.

A Taylor expansion of the so-called lens equation \eqref{eq:lenseq} yields the Jacobian of the $\vec{\theta} \rightarrow \vec{\beta}$ mapping, which defines the magnification tensor \citep[e.g.][]{BS01}
\begin{equation}\label{eq:jacobi}
  a_{ij}(\vec{\theta}) = \frac{\partial \vec{\beta}}{\partial \vec{\theta}}= \left( \delta_{ij} - \phi_{,ij}\right) \equiv \left(\begin{array}{cc} 1 - \kappa - \gamma_1 & -\gamma_2 \\ -\gamma_2 & 1 - \kappa + \gamma_1\end{array}\right)\;,
\end{equation}
where $\delta_{ij}$ is the Kronecker symbol, and the two components $\gamma_{1/2}$ of the complex spin-2 shear have been introduced. Note that subscripts following a comma denote partial derivatives along that coordinate. Both shear and convergence are first derivatives of the deflection field $\vec{\alpha}$ (or second derivatives of the lensing potential)
\begin{eqnarray}\label{eq:kg2psi}
  \kappa  & = &\frac{1}{2} ( \alpha_{1,1} + \alpha_{2,2}) \,,\\
  \gamma_1 &= &\frac{1}{2} ( \alpha_{1,1} - \alpha_{2,2})\,,\\
  \gamma_2 &= & \alpha_{1,2} = \alpha_{2,1} \,.
\end{eqnarray}

Therefore, starting from pixelised maps of the deflection field $\alpha_{1/2}(i,j)$ in a thin slice of the lightcone, one can easily derive $\gamma_{1/2}(i,j)$ and $\kappa(i,j)$ with finite differences or Fast Fourier Transforms (FFTs), even if $\alpha$ is only known on a finite aperture, without periodic boundary conditions.
Conversely, starting from a convergence map $\kappa(i,j)$, it is impossible to integrate \eqref{eq:poisson} with FFTs to get $\alpha$ (and then differentiate again to get $\gamma$) without introducing edge effects, if periodic boundary conditions are not satisfied. 

Additionally, we also introduce the scalar magnification $\mu$ which is the inverse determinant of the magnification tensor $a_{i,j}$ of Eq.~\eqref{eq:jacobi}.

\subsection{Propagation of rays in a continuous lumpy Universe}

On cosmological scales, light rays cross many over/under-dense extended regions at different locations. Therefore, the thin lens approximation does not hold.
The transverse deflection induced by an infinitely thin lens plane is still given by the above equations but one needs to fully integrate the trajectory of rays along their path. Therefore, for a given source plane at comoving distance $\chi_s$, the source plane position of a ray, initially observed at position $\vec{\theta}$ is given by the continuous implicit (Voltera) integral equation \citep{Jain1997}:
\begin{equation}\label{eq:fullpropagation}
\vec{\beta}(\vec{\theta}, \chi_{\rm s}) = \vec{\theta} -\frac{2}{c^2} \int_0^{\chi_{\rm s}} \der\chi\: \frac{\chi_{\rm s}-\chi}{\chi_{\rm s} \ \chi} \vec{\nabla}_{\beta} \phi \left(\vec{\beta}(\vec{\theta},\chi), \chi\right) \,.
\end{equation}

To first order, one can evaluate the gravitational potential along an unperturbed path, so that:
\begin{equation}\label{eq:Bornpropagation}
\vec{\beta}(\vec{\theta}, \chi_{\rm s}) = \vec{\theta} -\frac{2}{c^2} \int_0^{\chi_{\rm s}} \der\chi\: \frac{\chi_{\rm s}-\chi}{\chi_{\rm s} \ \chi} \vec{\nabla}_{\theta} \phi \left(\vec{\theta},\chi\right) \,.
\end{equation}
This is known as the \textit{Born approximation}, which is common in many diffusion problems of physics. 
An interesting property of the Born Approximation  is that the relation between $\vec{\beta}$ and $\vec{\alpha}$ can be reduced to an effective thin lens identical to \eqref{eq:lenseq} allowing the definition of an effective convergence, which is the divergence of the effective (curl-free) deflection field: $2 \kappa_{\rm eff} = \vec{\nabla}.\vec{\alpha}_{\rm eff}$.

When the approximation does not hold, the relation between $\vec{\beta}$ and $\vec{\alpha}$ can no longer be reduced to an effective potential and some curl-component may be generated, implying that the magnification tensor is no longer symmetric but requires the addition of a rotation term and so-called B-modes in the shear field. In this more general framework, the magnification tensor should be rewritten
\begin{equation}\label{eq:jacobiROT}
  a_{ij}(\vec{\theta}) = \left(\begin{array}{cc} 1 - \kappa - \gamma_1 & -\gamma_2 -\omega \\ -\gamma_2 + \omega & 1 - \kappa + \gamma_1\end{array}\right)\;.
\end{equation}
with the following definitions of the new lensing rotation term $\omega$ (and revised $\gamma_2$)
\begin{eqnarray}\label{eq:kg2aROT}
  \gamma_2 & = &\frac{1}{2} ( \alpha_{1,2} + \alpha_{2,1}) \,,\\
  \omega  & = &\frac{1}{2} ( \alpha_{1,2} - \alpha_{2,1}) \,.
\end{eqnarray}
The image plane positions where $\omega\ne 0$ are closely related to the lines of sight along which some substantial lens-lens coupling may have occurred.

\subsection{The multiple lens planes approximation}
The numerical transcription of equation~\eqref{eq:fullpropagation} in the \hagn~past line-cone requires the slicing of the latter into a series of parallel transverse planes, which could simply be the 22,000 slabs dumped by \ramses~at runtime every coarse time step. These are too numerous and can safely be stacked into thicker planes by packing together 40 consecutive slabs\footnote{This number was chosen as a tradeoff between the typical number of CPU cores in the servers  used to perform the calculations and the preservation of the line-of-sight native sampling of lightcone.}.  Here  500 slices of varying comoving thickness are produced  all the way to redshift $z=7$ to compute either the deflection field or the projected surface mass density as described below.

The discrete version of the equation of ray propagation \eqref{eq:fullpropagation} for a fiducial source plane corresponding to the distance of the plane $j+1$ reads:
\begin{equation}\label{eq:crude_multi_plane}
 \vec{\beta}^{j+1} = \vec{\theta} -\sum^{j}_{i=1} \frac{D_{i;j+1}}{D_{j+1}} \vec{\alpha}^i (\vec{\beta}^i)\,,
\end{equation} 
where $\vec{\alpha}^i $ is the deflection field in the lens plane $i$, $D_{j+1}$ is the angular diameter distance between the observer and the plane $j+1$, and $D_{i;j+1}$ the angular diameter distance between planes $i$ and $j+1$.
Therefore, as sketched in Fig.~\ref{fig:raytracing}, rays are recursively deflected one plane after the other, starting from unperturbed positions on a regular grid $\vec{\theta} \equiv \vec{\beta}^1$.

The practical implementation of the recursion in equation~\eqref{eq:crude_multi_plane} is computationally cumbersome and demanding in terms of memory because the computation of the source plane positions $\vec{\beta}^{j+1}$ requires holding all the $j$ previously computed source plane positions. Instead, this paper follows the approach of \citet{hil++09}, who showed that equation~\eqref{eq:crude_multi_plane} can be rewritten as a recursion over only three consecutive planes\footnote{This recursion requires the introduction of an artificial $\vec{\beta}^0\equiv\vec{\beta}^1 = \vec{\theta}$ slice in the initial setup.}
\begin{equation}\label{eq:multi_plan}
 \vec{\beta}^{j+1}= \left( 1 - \frac{D_{j}}{D_{j+1}} \frac{D_{j-1;j+1}}{D_{j-1;j}} \right)  \vec{\beta}^{j-1} +  \frac{D_{j}}{D_{j+1}} \frac{D_{j-1;j+1}}{D_{j-1;j}}  \vec{\beta}^{j}  -  \frac{D_{j;j+1}}{D_{j}}  \vec{\alpha}^{j} (\vec{\beta}^{j} )\,.
\end{equation}

\begin{figure}[h!]
\includegraphics[width=0.49\textwidth]{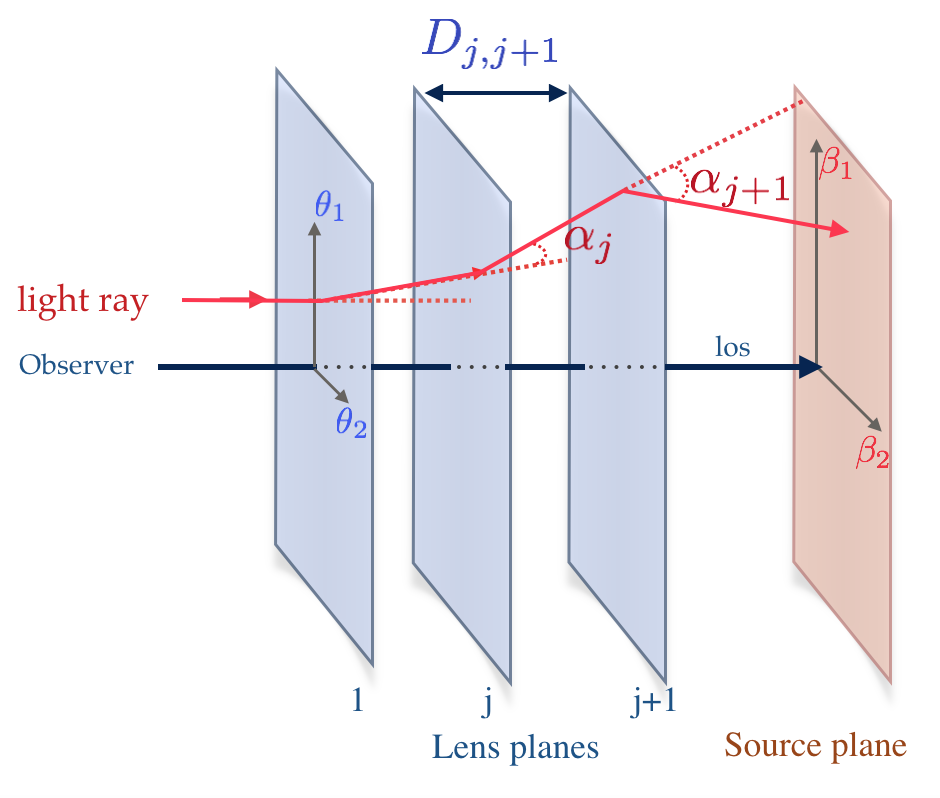}  
\caption{Schematic view of the propagation of a light ray through a lightcone sliced into multiple discrete lens planes. The ray  (red line) is deflected at each intersection with a thin lens plane. The deflection field is defined for each plane depending of the angular position on this plane $\vec{\alpha}^{j}(\vec{\beta}^{j})$.}
\label{fig:raytracing}
\end{figure}

Besides this thorough propagation of light rays 
source plane positions and associated quantities (convergence $\kappa$, shear $\gamma$, rotation $\omega$) are additionally computed  using the Born approximation, following the discrete version of equation~\eqref{eq:Bornpropagation}:
\begin{equation}\label{eq:multi_plan_born}
 \vec{\beta}^{j+1} = \vec{\theta} -\sum^{j}_{i=1} \frac{D_{i;j+1}}{D_{j+1}} \vec{\alpha}^i (\vec{\theta})\,
\end{equation} 

The deflection maps in each lens plane are computed on a very fine grid of pixels of constant angular size. In order to preserve the $\sim 1 \kpc$ spatial resolution allowed by the simulation at high redshift,  $36,000\times 36,000$ deflection maps are built in the narrow 1 deg lightcone. The deflection maps in the low redshift 2.25 sq deg wide cone reaching $z=1$ are computed on a coarser $20,000\times20,000$ pixels grid since the actual physical resolution of the simulation at low redshift does justify the $0.1 \textrm{ arcsec}$ resolution of the narrow 1 sq deg field-of-view. Even though the image plane positions $\vec{\theta}=\vec{\beta}^1$ are placed on the regular pixel grid, the deflections they experience must be interpolated in between the nodes of the regular deflection map as they progress backward to a given source plane. This is done with a simple bilinear interpolation scheme.

\subsection{Total deflections from the \ramses~accelerations}\label{ssec:OBB}
Let us now describe how to obtain $\alpha$ to use it in Eqs.~\eqref{eq:multi_plan} and \eqref{eq:multi_plan_born}.
The first method uses the gravitational acceleration field which is registered on each (possibly-refined) grid location inside the lightcone. The very same gravitational field that was used to move particles and evolve Eulerian quantities in \ramses~is interpolated at every cell position and is therefore used to consistently derive the deflection field. The merits of the complex three-dimensional multi-resolution Poisson solver are therefore preserved and the transverse components of the acceleration fields can readily be used to infer the deflection field.
By integrating the transverse component of the acceleration along the light of sight, one can compute the deflection field according to equation~\eqref{eq:alphadef}.

To do so,  for each light ray, gas cells which intersect the ray are considered, and  the intersection length along the line-of-sight $l_i$ is computed. 
 Knowing the cell size $\delta_i$, and its orientation with respect to the line of sight, $l_i$ is deduced with a simple Oriented-Box-Boundary (OBB) algorithm \citep[e.g.][]{RTR3} in which it is assumed that all cells share the same orientation (flat sky approximation) and factorise out expensive dot products between normals to cell edges and the line of sight.
\begin{equation}
 \vec{\alpha} (\vec{\theta})= \frac{2}{c^2} \sum_{i \in \mathcal{V}(\vec{\theta})} \vec{\nabla_{\perp}}\phi_i (\vec{\theta}) \ l_i\;,
\end{equation}
where $\mathcal{V}(\vec{\theta})$ denotes the projected vicinity of a sky position $\vec{\theta}$.
 As shown in fig \ref{fig:tetris}, a fiducial light ray is drawn: at each lens plane, the deviation of the light is calculated as the direct sum of the transverse acceleration components recorded on the cells $i$, weighted by the intersection length $l_i$.
Here, the field of view is small and one can safely assume that light rays share the same orientation (flat sky approximation) and are parallel to the line-of-sight.

This method has the main advantage of preserving the gravitational force that was used when evolving the simulation. In particular, the way shot noise is smoothed out in the simulation to recover the acceleration field from a mixture of Lagrangian particles and Eulerian gas cells is faithfully respected in the raytracing.
In other word, the force felt by photons is very similar to the one felt by particles in the simulation.
Dealing with acceleration is also local, in the sense that the deflection experienced by a light ray (and related derivatives leading to e.g. shear and convergence) depends only on the acceleration of cells this ray crosses. The mass distribution outside the lightcone is therefore consistently taken into account via the acceleration field.

However, this method is sensitive to small artefacts which are present at the lightcone generation stage (i.e. simulation runtime) and which  could not be corrected without a prohibitive post-processing of the lightcone outputs. When the simulation dumps two given neighbouring slabs at two consecutive time steps, problems can happen if cells on the boundary between the two slabs have been (de-)refined in the mean time. As illustrated in Fig.~\ref{fig:tetris}, such cells can be counted twice or can be missing, if they are refined (or derefined) at the next time step. Those bumps and dips in the deflection map translate into saw-tooth patterns in the convergence maps. They are however quite scarce and of very modest amplitude. 

\begin{figure}[h!]
\includegraphics[width=0.49\textwidth]{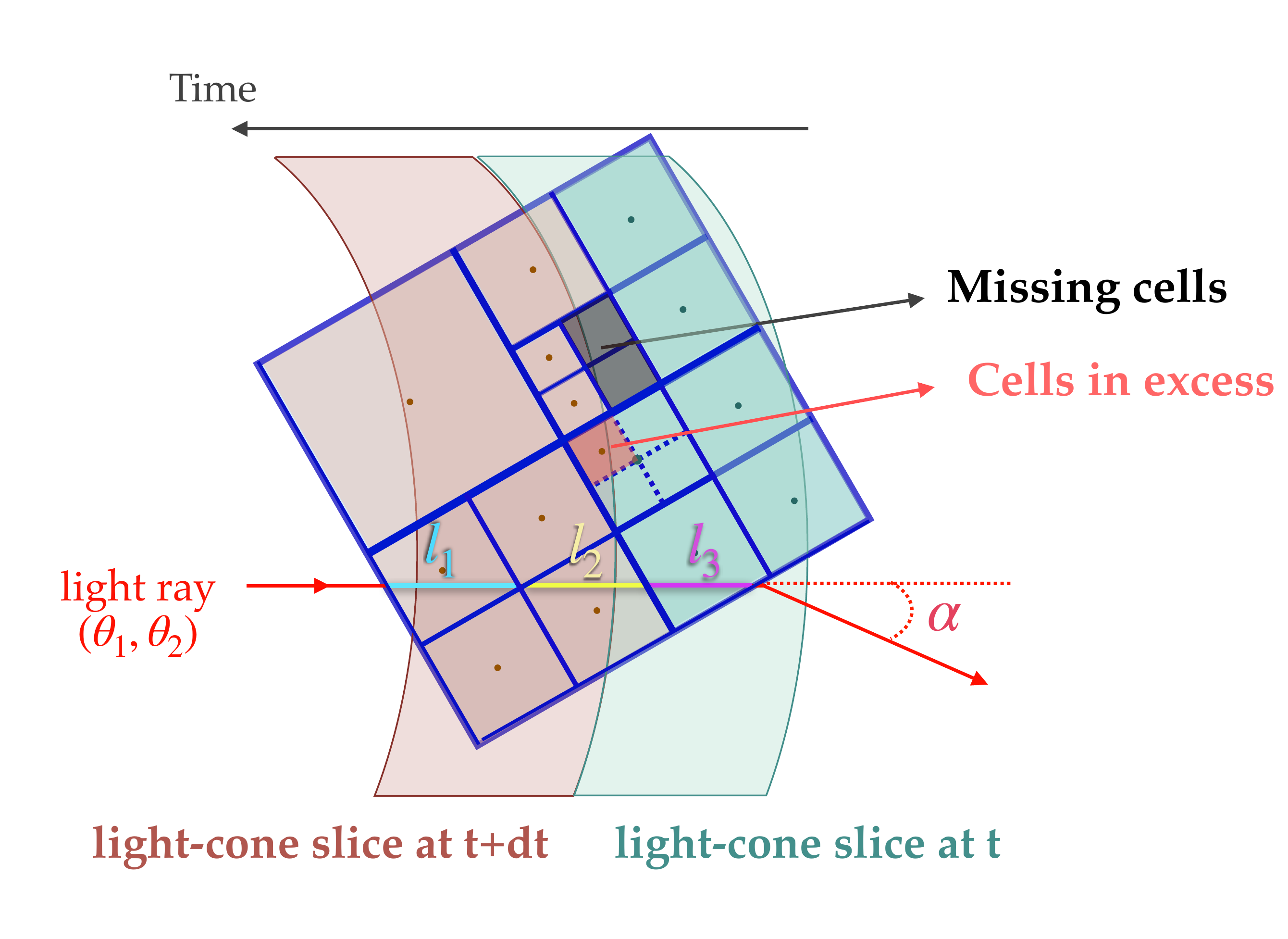}
\caption{Schematic view of the problem induced by cells at the boundary of slabs $j$ and $j+1$, which get refined between time $t$ and $t+\der t$. Missing cells (devoid of dots) or cells in excess (overlapping "dotted" cells of different colour) can end up as lightcone particles. A fiducial light ray is drawn to illustrate the intersection length $l_i$ between the ray and RAMSES cells.} 
\label{fig:tetris}
\end{figure}

A 100 arcsec wide zoom into the convergence map obtained with this method is shown in the left panel of Fig.~\ref{fig:metcomp}. The source redshift is $z_{\rm s}=0.8$. A few subdominant artefacts due to missing acceleration cells are spotted. 
They induce small correlations on scales smaller than a few arcsec and are otherwise completely negligible for our cosmological applications.

\begin{figure*}[h!]
\includegraphics[width=\textwidth]{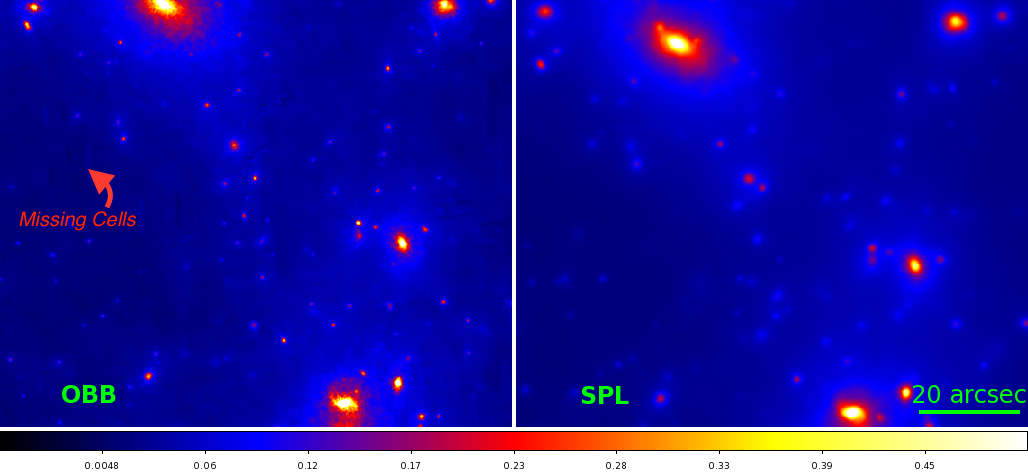}  
\caption{Comparison of $z_{\rm s}=0.8$ convergence maps obtained with the OBB method (integration of transverse accelerations in cells, {\it left}) and with the SPL method (projection of particles onto convergence planes after adaptative Gaussian smoothing, {\it right}). The latter method applies a more aggressive smoothing which better erases shot noise. Inaccuracies of long range deflections in the SPL method due to edge effects translate into a global shift for some galaxies, as compared to OBB. With this method, some missing acceleration cells produce modest artefacts on small scale, here and there.}
\label{fig:metcomp}
\end{figure*}

\subsection{Projection of smoothed particle density}\label{ssec:SPL}
The second method of computing the deflection maps in thin lens planes is more classical: it relies on the projection of particles onto surface density maps which are then turned into deflection maps. If the line-of-sight integration is performed under the Born approximation, the  Fourier inversion going from the projected density to the deflection is just done once starting from the effective convergence. Otherwise, with the full propagation, many FFTs inversions on projected density maps that do not fulfil the periodic boundary condition criterion imply an accumulation of the inaccuracies in the Fourier inversion.

First of all, this method allows us to separate the contribution of each matter component to the total deflection field. One can therefore compute the contribution of stars or gas to the overall lensing near a given deflector, something that is not possible with the acceleration method since only the total acceleration is computed by the simulation.

In addition, one can project particles with an efficient and adaptive smoothing scheme. Instead of a standard nearest grid point or cloud-in-cell projection, a gaussian filter (truncated at 4 times the standard deviation $\sigma$) is used in which the width of the smoothing filter $\sigma$ is tuned to the local density, hence following the Smooth Particle Lensing (SPL) method of \citet{Aubert++07}. Since the AMR grid of \ramses~is adaptive, the resolution level around a given particle position from the neighbouring gas cells can  be recovered. This thus bypasses the time consuming step of building a tree in the distribution of particles, which is at the heart of the SPL method.

To illustrate the merits of this method and for comparison with the previous one, let us show the same region of simulated convergence fields for a source redshift $z_{\rm s}=0.8$ in the right panel of Fig.~\ref{fig:metcomp}.
This adaptive gaussian smoothing (referred to as SPL method below) seems more efficient at smoothing the particle noise out. Between the two methods, we notice small displacements of some galaxies of a few arcsec. They are due to the long range inaccuracies generated by the Fourier inversions. 

\subsection{Lensing of galaxy and halo catalogues}\label{ssec:defl-points}
In order to correlate galaxies (or halos) in the lightcone with the convergence or shear field around them and, hence, measure their GGL, one has to shift their  catalogue positions $\vec{\beta}$ (which are intrinsic source plane coordinates) and infer their observed lensed image plane positions $\vec{\theta}$. These are related by the thorough lens equation \eqref{eq:fullpropagation}, or its numerical translation \eqref{eq:crude_multi_plane}. However, this equation is explicit for the $\theta\rightarrow \beta$ mapping, only. The inverse relation, which can be multivalued when strong lensing occurs, has to be solved numerically by testing for every image plane mesh $\vec{\theta}_{ij}$ whether it surrounds the coordinates $\vec{\beta}^{\rm gal}$ of the deflected galaxy  when cast into the source plane $\vec{\beta}_{ij}$ \citep[e.g.][]{SEF92,keeton01soft1,bartelmann03c}. Because the method should work in the strong lensing regime, regular rectangular meshes may no longer remain convex in the source plane and, therefore, it is preferable to split each mesh into two triangles. Those triangles will map into triangles in the source plane and one can safely test whether $\vec{\beta}^{\rm gal}$ is inside them. In order to speed up the test on our large pixel grids, the image plane is partitioned into a quad-tree structure that recursively explore finer and finer meshes. The method is actually very fast and yields all the image plane antecedents of a given galaxy position $\vec{\beta}^{\rm gal}$. This provides us the updated catalogues of halos and galaxies.

Obviously, when measuring the GGL signal in the Born approximation, catalogue entries do not need to be deflected and therefore source plane and image plane coordinates are identical.

\subsection{Summary of generated deflection maps}\label{ssec:subsum}

Table \ref{tab:pros_cons} summarises the main advantages and drawbacks of the OBB and SPL methods.

Altogether,   $2 \times2 $ (OBB/SPL and Born approximation/full propagation) deflection maps were generated for each of the 246 source planes all the way to $z=1$ in the wide opening angle field. Likewise, we obtained $2 \times 2 $ maps for each of the 500 source planes all the way to $z=7$ in the narrow opening angle field.

\begin{table*}{
\begin{tabular}{C{3.5cm}  C{7cm}  C{7cm} }
\hline
   & OBB & SPL \\
    \hline\hline
    Deflection (per plane) & integration of transverse acceleration &  particles adaptively smoothed and projected\\
    && onto density planes \\
Large scale  &  matter outside the lightcone is taken into account  & Edge effects due to Fourier Transforms  \\ 
Small scale & uses the multi-scale \ramses~potential & smoothing reduces small-scale features\\
Cells missing/in excess &  produces small scale artefacts & unaffected  \\
Matter component  & only for the total matter  & can individually consider DM, stars, and gas \\
   \hline\hline
\end{tabular}
\caption{Summary of the main properties of the SPL and OBB methods ray-tracing methods.}
\label{tab:pros_cons}}
\end{table*}

\section{Cosmic shear}\label{sec:cosmicshear}
This section  assesses the validity of our ray-tracing methods by measuring 1-point and 2-point statistics of the lensing quantities like convergence, and (reduced-)shear. It also compares those finding with other methods.

The focus is on the impact of baryons on small scales for multipoles $\ell\gtrsim 2000$ to check whether the baryonic component couples to other non-linear effects like the shear -- reduced shear correction and Beyond-Born treatments.

\subsection{Convergence 1-point statistics}

The most basic quantity that one can derive from the convergence field shown in the right panel of Fig.~\ref{fig:convergence_map} is the probability distribution function (PDF) of the convergence. The Fig.~\ref{fig:convergence_map} shows this quantity which is extremely non-Gaussian at the $\sim 1\arcsec$ resolution of the map. One can see the skewness of the field with a prominent high-end tail and a sharp fall off of negative convergence values.

\begin{figure*}[h]
\begin{minipage}{0.7\textwidth}
	\includegraphics[width=0.8\textwidth]{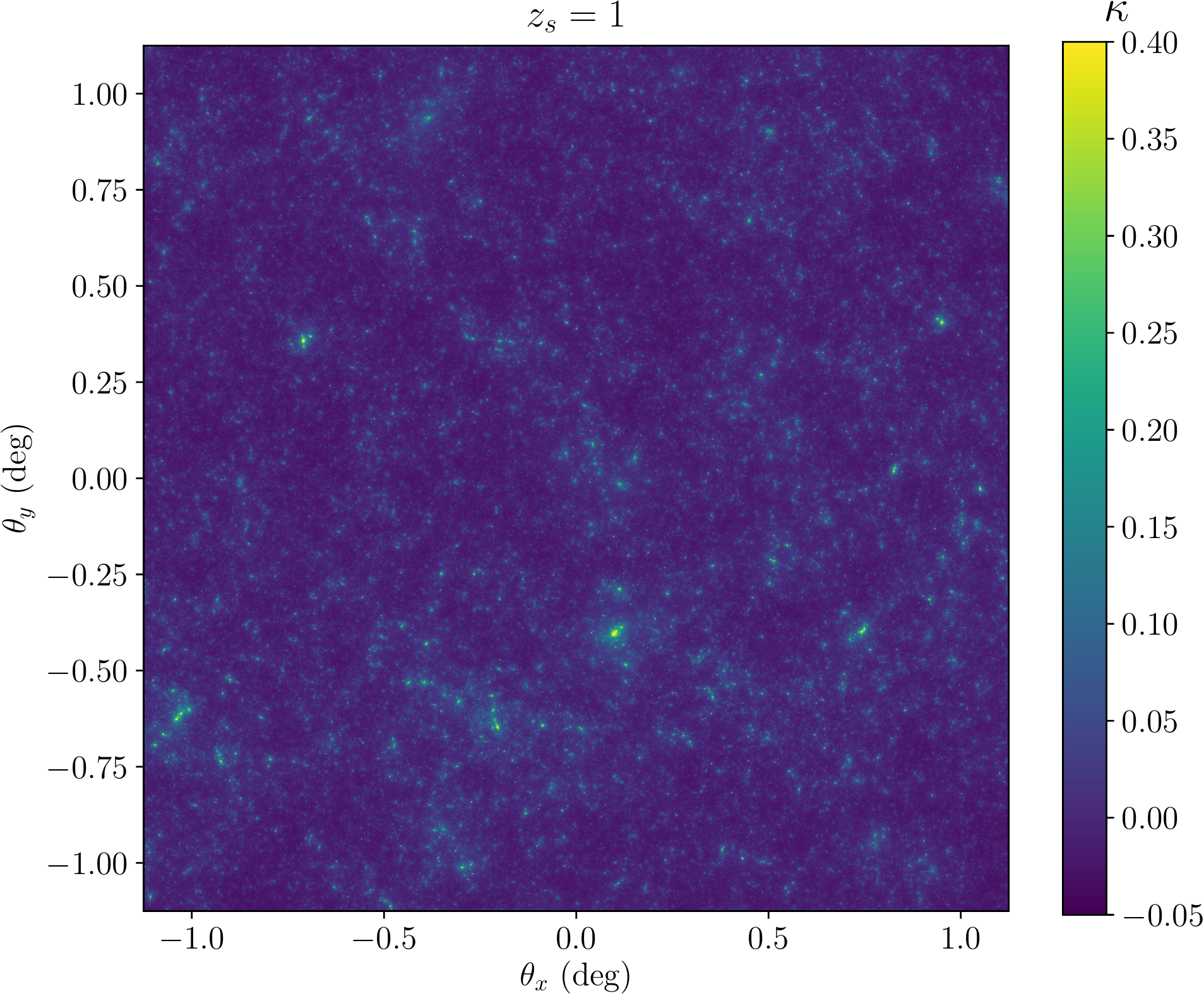}  
\end{minipage}
\begin{minipage}{0.67\textwidth}
	\hspace*{-0.85cm} \includegraphics[width=0.4\textwidth]{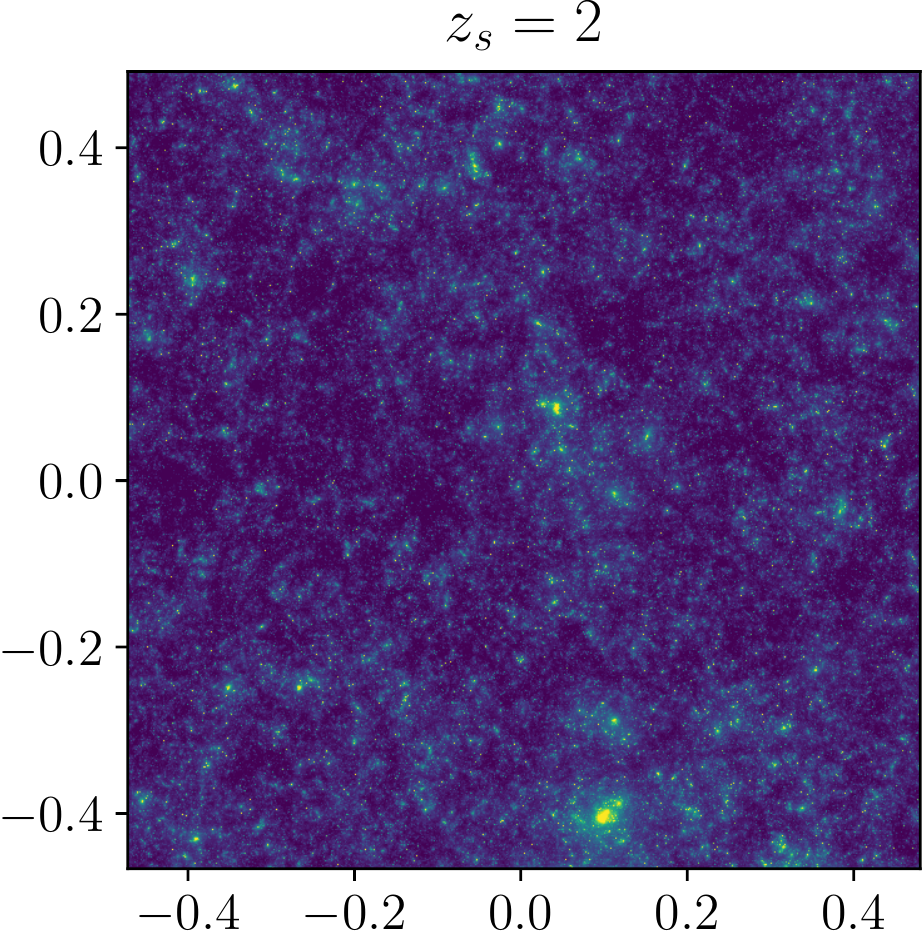}  \\
	\hspace*{-2.cm} \includegraphics[width=0.6\textwidth]{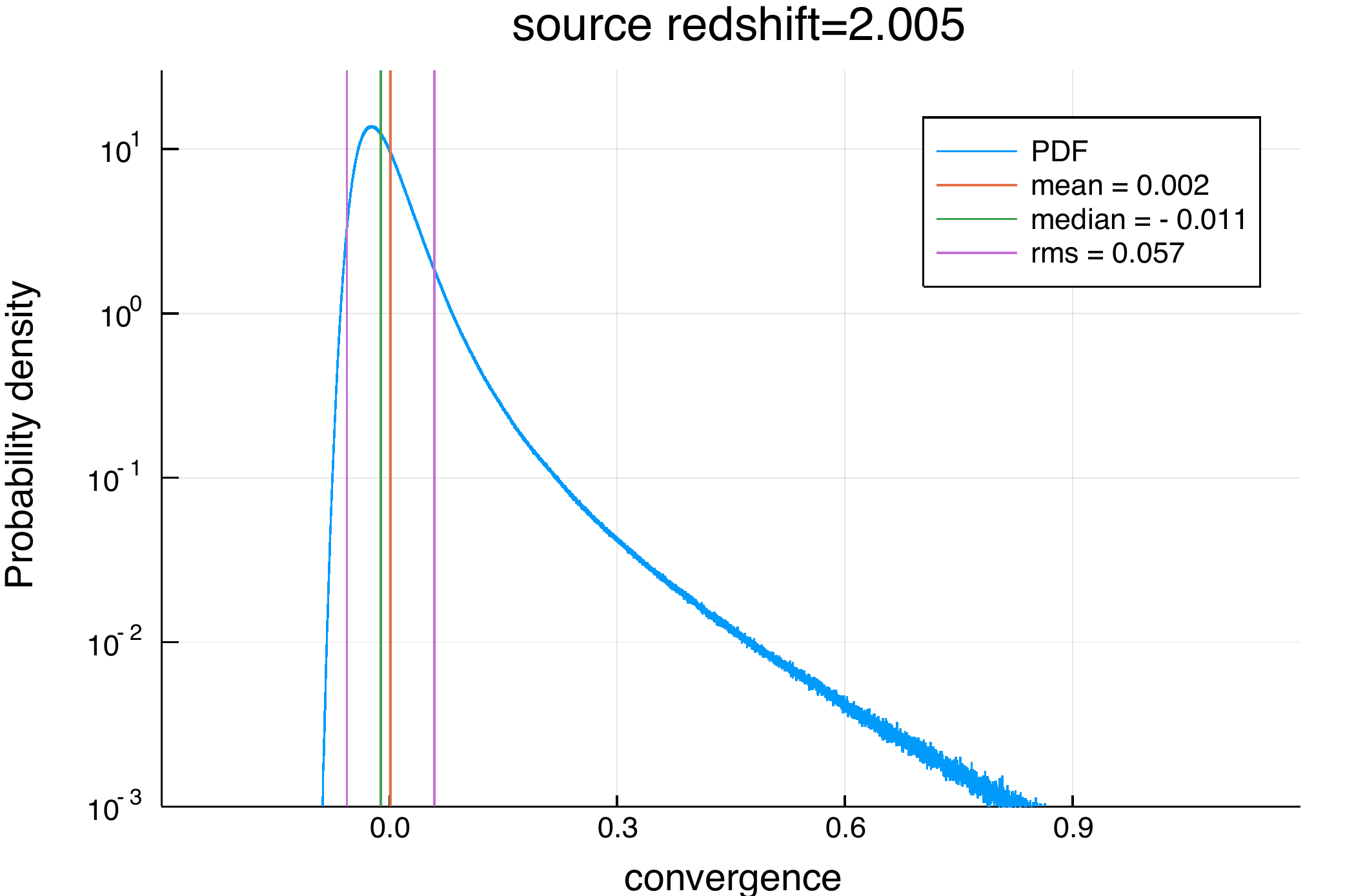}  
\end{minipage}

\caption{ {\it Left panel:} Convergence map generated with a $0\farcs1$ pixel grid over a $2.25 \times 2.25$ square degrees field of view for a fiducial source plane at $z_{\rm s}\sim~1$.
 {\it Right panel:} Convergence map with a field of view of $1$ sq. deg. at $z_s \sim 2$, and its corresponding convergence PDF showing the characteristic skewed distribution.
} 
\label{fig:convergence_map}
\end{figure*}

\subsection{Convergence power spectrum}
 In Fourier space, the statistical properties of the convergence field are commonly characterised by its angular power spectrum $P_\kappa (l)$,
\begin{equation}
\langle  \hat{\kappa}(\vec{\ell}) \, \hat{\kappa}^*(\vec{\ell'}) \rangle  = (2 \pi)^2\, \delta_{D}(\vec{\ell}-\vec{\ell'})\, P_{\kappa}(\ell)\,.
\end{equation}
where $\delta_{D}(\vec{\ell}$ is the Dirac delta function.
For two fiducial source redshifts ($z_{\rm s}=0.5$ and $z_{\rm s}=1$), Fig.~\ref{fig:pkappa} shows the angular power spectrum of the convergence obtained with the two ray tracing techniques: the OBB and SPL methods (respectively solid magenta and solid cyan curves). The low redshift ones are based on the $2.25\,{\rm deg}$ wide lightcone. They are thus more accurate on larger scales $\ell \lesssim 10^3$, even though the large sample variance will not permit quantitative statements. 
On small scales ($\ell \sim 2\times 10^5$), the additional amount of smoothing implied by the SPL projection of particles onto the lens planes induces a deficit of power with respect to the less agressive softening of the OBB method in which shot noise has not been entirely suppressed (see Fig.~\ref{fig:metcomp}).

The middle panel of Fig.~\ref{fig:pkappa} shows the difference between power spectra inferred using the Born approximation or with the full multiple lens plane approach for the OBB method.
For angular scales $ \ell \lesssim 8 \times 10^4$, we find differences between the two propagation methods that are less than 0.5\%, or so, which is totally negligible given possible numerical errors and sampling variance limitations.
 At lower angular scales $ \ell \gtrsim 10^5$, departures rise above the few percent level. Note that this scale also corresponds to scale where shot noise (from DM particles) and convergence power spectral are of equal amplitude (yellow shaded area). Below these very small scales, close to the strong lens regime, the Born approximation may start to break down \citep{Schafer2012}.


Under the Limber and Born approximations, one can express the convergence power spectrum as an integral of the three-dimensional non-linear matter power spectrum $P_\delta$ \citep{Limber,blandford91,miralda91,K92} from the observer to the source plane redshift or corresponding co-moving distance $\chi_{\rm s}$:
\begin{equation}\label{eq:limber}
P_{\kappa}(\ell) = \left( \frac{3 \Omega_{\rm m} H_0^2 }{2c^2} \right)^2 \int_0^{\chi_{\rm s}}  \der\chi \left( \frac{\chi (\chi_{\rm s} - \chi)}{\chi_{\rm s} a(\chi)}\right)^2   P_{\delta}\left(\frac{\ell}{\chi},\chi \right) \,,
\end{equation}
where $a$ is the scale factor and where no spatial curvature of the Universe was assumed for conciseness and because the cosmological model in \hagn~is flat.
As a validation test of our light deflection recipes, the lensing power spectrum derived from the actual ray-tracing is compared to  an integration of the three-dimensional matter power spectrum measured by \citet{Chisari++18} in the \hagn~simulation box. The red curve is the direct integration of $P_\delta(k)$ power spectra and the dashed parts of the lines corresponds to a power-law extrapolation of the $P_\delta(k)$ down to smaller scales. 
In the range $3\,000 \lesssim \ell \lesssim 3\times 10^5$, an excellent agreement is found between the red curve and the spectra inferred with our two ray-tracing techniques. On larger scales, the cosmic variance (which is different in the full simulation box and the intercept of the box with the lightcone) prevents any further agreement. This is also the case for $\ell \gtrsim 3\times 10^5$ where some possibly left over shot noise in the raytracing maps and the hazardous high-$\ell$ extrapolation of the three-dimensional power spectra complicate the comparison. In addition, the low-$\ell$ oscillations of the spectrum is likely to originate from the replicates of the simulation box throughout the past lightcone.

\citet{Chisari++18} also measured matter power spectra in the Horizon-DM simulation at various redshifts. This simulation is identical to \hagn~in terms of initial conditions but has been run without any baryonic physics in it after having rescaled the mass of DM particles to conserve the same total matter density \citep{Peirani++17,Chisari++18}. The integration of this DM-only power spectrum allows to get a sense on the effect of baryons in the DM-distribution itself.
Just like the red curve was showing the result of the Limber integral in equation~\eqref{eq:limber} for \hagn, the dark blue curve shows the same integral for Horizon-DM. The latter has much less power for $\ell \gtrsim  2\times 10^4$ than either the integration of the full physics \hagn~matter power spectrum (red) or that derived directly from ray-tracing (purple or green). The boost of spectral amplitude is due to cool baryons in the form of stars at the center of halos. Moreover, we notice a deficit of power on scales $2\times 10^3 \lesssim \ell \lesssim 2\times 10^4$ for the full physics simulation. As pointed out by \citet{Semboloni++11}, the pressure acting on baryons prevents them from falling onto halos as efficiently as dark matter particles, hence reducing the depth of the potential wells, when compared to a dark-matter only run. This effect has already been investigated with more sensitivity on the three-dimensional matter power spectrum in the Horizon-AGN simulation \citep{Chisari++18}, and a clear dip in the matter density power spectrum of the full physics simulation is observed on scales $1 \lesssim k \lesssim 10 \, h \Mpc^{-1}$. Here, the projection somewhat smears out this dip over a larger range of scales but   a $\sim 15\%$ decrease in amplitude is typically observed for $\ell=10^4$ at $z_{\rm s}=0.5$. In order to better see the changes due to the inclusion of the baryonic component, we traced rays through the lightcone by considering only the DM particles of the \hagn~run with the SPL method. For this particular integration of rays trajectories, we multiplied the mass of the dark matter particules by a factor $1+\Omega_{\rm b}/\Omega_{\rm DM}$ (where $\Omega_{\rm DM}=\Omega_{\rm m}-\Omega_{\rm b}$) to get the same overall cosmic mean matter density. The cyan curve in the upper panel shows the resulting convergence power spectrum. 
The ratios between the total full physics convergence power spectrum and the rescaled dark matter contribution of this power spectrum at $z_{\rm s}=0.5, 1.0$ and $1.5$ are shown in the bottom panel and further illustrates the two different effects of baryons on intermediate and small scales.

By considering two raytracing methods to derive the convergence power spectrum, and by asserting that consistent results are obtained by integrating the three-dimensional matter power spectrum, let us now look for small scale effects involving the possible coupling between the baryonic component and shear -- reduced shear corrections.

\begin{figure}[h]
\includegraphics[width=0.49\textwidth]{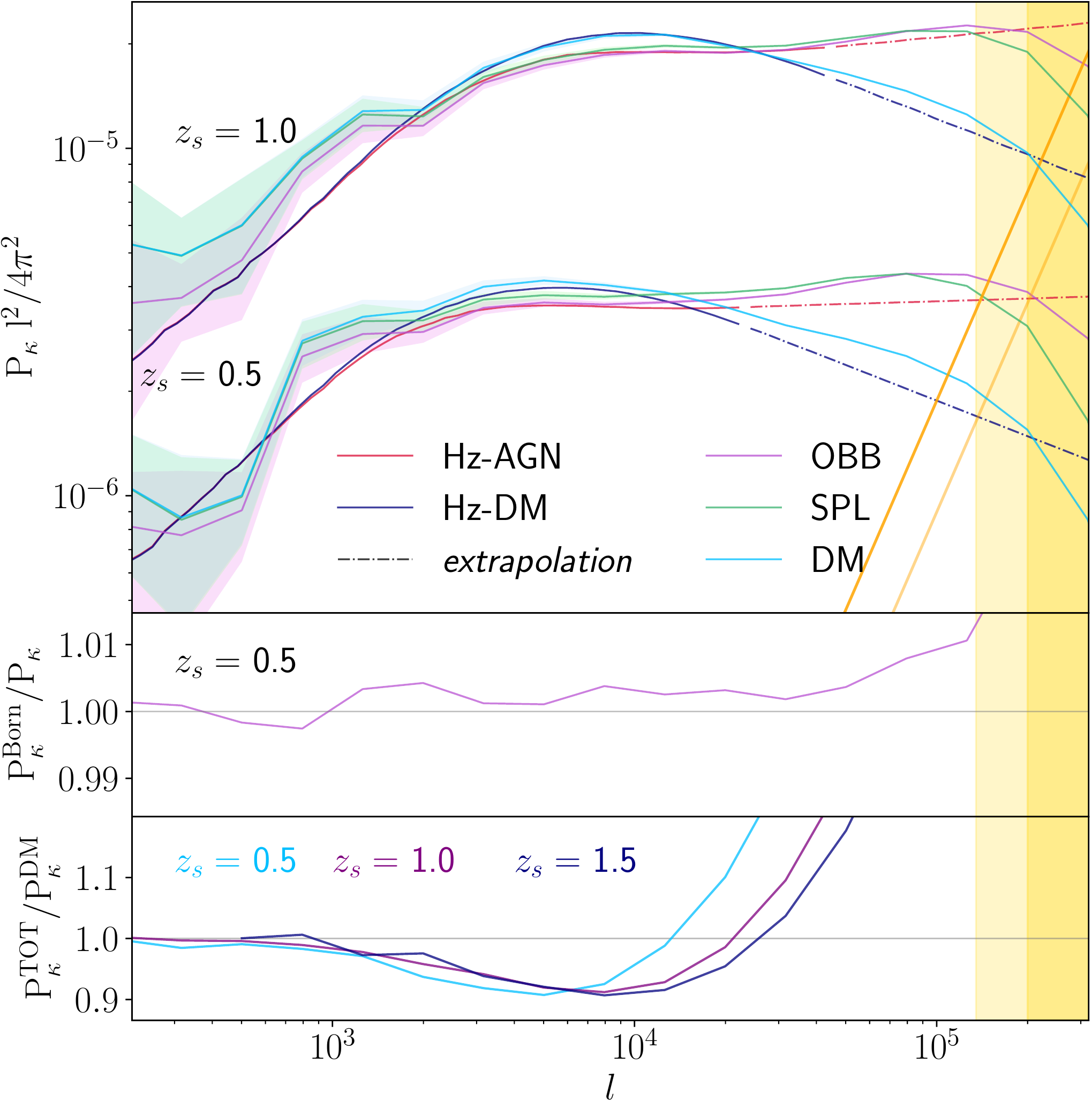}
\caption{{\it Upper panel:} Convergence power spectra for source redshift $z_{\rm s}=1$ (top) and $z_{\rm s}=0.5$ (bottom) derived with the OBB (magenta) and the SPL (green) methods. The more aggressive smoothing of this latter method translates into a faster high-$\ell$ fall-off. The cyan curves (DM) only account for the dark matter component (rescaled by $1+\Omega_{\rm b}/ \Omega_{\rm M}$). The red curve corresponds to the direct integration of the 3D total matter power spectrum (Limber approximation)  in the Horizon-AGN simulation  (Hz-AGN). The blue curves is the direct integration of the Horizon-DM (dark matter only) matter power spectrum (Hz-DM). Dashes reflect regimes where the 3D spectra of \citet{Chisari++18} was extrapolated by a simple power law (extrapolation). 
The yellow lines show the particle shot-noise contribution at two different redshifts.
{\it Middle panel:} ratio of the $z_{\rm s}=0.5$ convergence power spectra obtained with the Born approximation and the proper multiple lens plane integration showing only very small changes up to $\ell\sim 10^5$. {\it Bottom panel:} ratio of the dark-matter only to total convergence power spectra at $z_{\rm s}=0.5, \ 1.0$ and $1.5$ for the SPL method.}
\label{fig:pkappa}
\end{figure}

\subsection{Shear -- reduced shear corrections to 2-point  functions}
In practical situations, rather than the convergence power spectrum, which is not directly observable, wide fields surveys give access to the angular correlation of pairs of galaxy ellipticities. The complex ellipticity\footnote{$\varepsilon= (a-b)/(a+b) {\rm e}^{2i \varphi}$, with $a$ and $b$, respectively, the major and minor axis of a given galaxy, and $\varphi$ is the orientation of the major axis.} $\varepsilon$ is directly related to the shear $\gamma$. The relation between the ensemble mean ellipticity and the shear is in fact
\begin{equation}
  \langle \varepsilon \rangle = g \equiv \frac{\gamma}{1-\kappa} \simeq \gamma \,,
\end{equation}
with, $g$, the so-called reduced shear. Therefore, the two point correlations of ellipticities  and shear only match when the convergence $\kappa$ is small. Since the regions of large convergence are typically the centres of halos where the contribution of cooled baryons is highest, one might expect a coupling between the inclusion of baryons and the shear reduced-shear corrections needed to properly interpret the cosmological signal carried by the 2-point statistics \citep[e.g.][]{White05,Kilbinger10}

Owing to the spin-2 nature of ellipticity, one can define the angular correlation functions $\xi_\pm$
\begin{eqnarray}
 \xi_\pm(\theta) &= &\langle \gamma_+(\vartheta+\theta) \gamma_+(\vartheta) \rangle_\vartheta \pm \langle \gamma_\times(\vartheta+\theta) \gamma_\times(\vartheta) \rangle_\vartheta \label{eq:xipmdef1} \,,\\
   & =& 2 \pi \int \der\ell\, \ell J_{0/4}(\theta \ell) P_\kappa(\ell) \label{eq:xipmdef2} \,,
\end{eqnarray}
where $\gamma_+$ and $\gamma_\times$ are defined with respect to the separation vector between two galaxies or, here any two image plane positions at separation $\theta$. $J_0$ and $J_4$ are 0th and 4th order Bessel functions.

Instead of the shear, observers can only measure associated ellipticities $\epsilon$, which should thus replace $\gamma$ in equation~\eqref{eq:xipmdef1} in practical measurements.  
The reduced shear maps were computed together with shear and convergence maps, so as to measure the modified $\xi_+$ and $\xi_-$ angular correlations to compare them with the actual correlation functions. For efficiency, the {\tt Athena} code\footnote{\url{http://www.cosmostat.org/software/athena}} was used to compute correlation functions.

The results are shown in Fig.~\ref{fig:xipm} for a fiducial source redshift $z_{\rm s}=0.5$. Here $\xi_+^g$ and $\xi_+^\gamma$ only depart from one another at the $\sim 2-3\%$ level on angular separations $\sim 1\arcmin$. The effect is slightly stronger for $\xi_-$ which is known to be more sensitive to smaller non-linear scales than $\xi_+$, but also more difficult to measure in the data because of its lower amplitude. On $1\arcmin$ scales,  $\xi_-^g/\xi_-^\gamma -1 \simeq 7-8\%$.
Like for the power spectra in the previous subsection, the cyan curves represent the correlations $\xi_\pm^\gamma$ for the rescaled DM contribution. The bottom panel shows the ratio of rescaled DM over full physics reduced shear correlation functions, further illustrating the effect of baryons on small scales. Again, $\xi_-$ responds more substantially to the inclusion of baryons. The deficit of correlation amplitude when baryons are taken into account peaks at $3-4\arcmin$ and is of order 10\%. Below $1\arcmin$, the effect starts to increase but those scales are never used in practical cosmic shear applications.

\begin{figure}[h]
\includegraphics[width=0.49\textwidth]{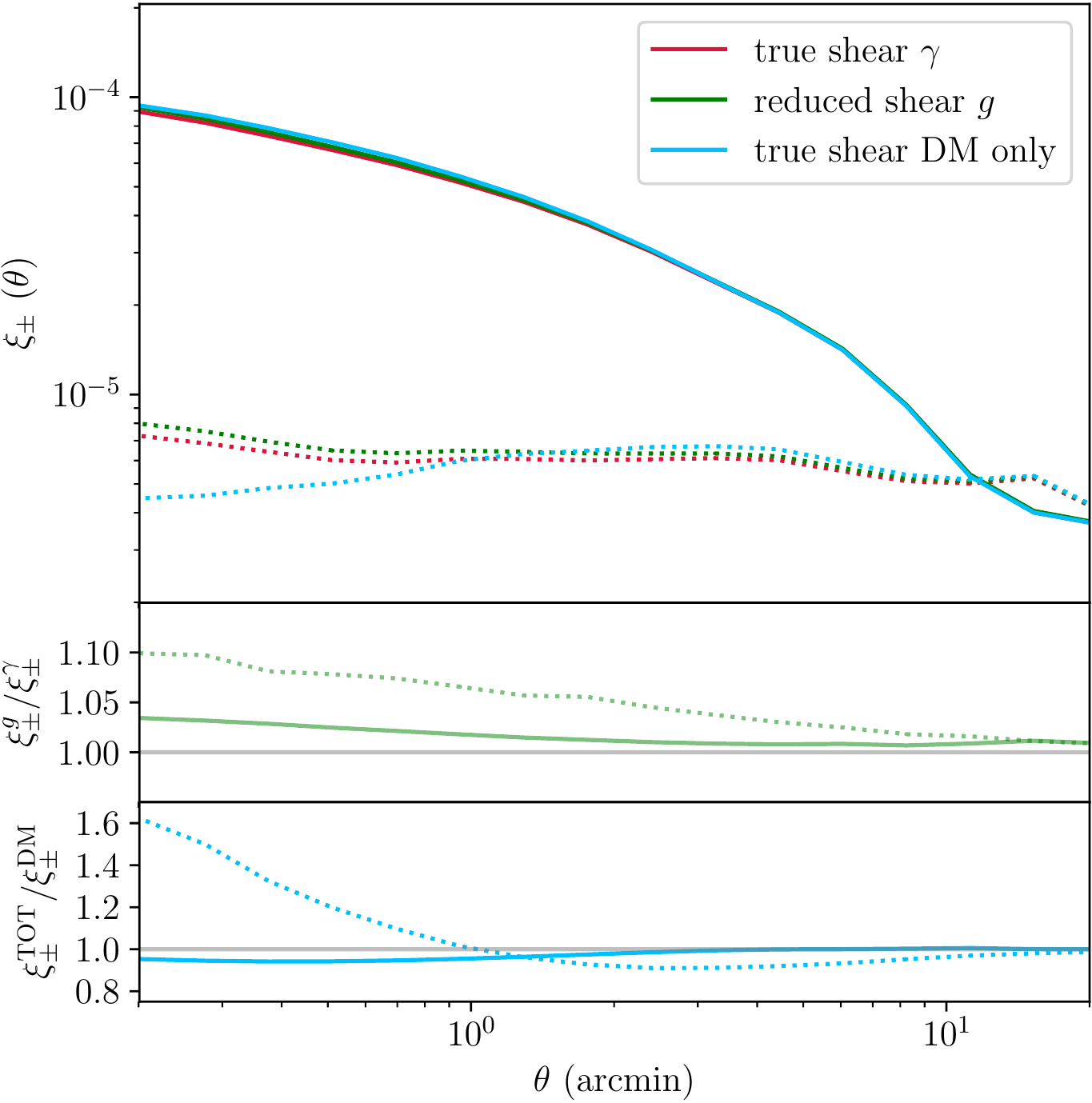}  
\caption{{\it Upper panel:} Two-point shear correlation functions $\xi_+$ (solid lines) and $\xi_-$ (dotted lines) for a fiducial source redshift $z_{\rm s}=0.5$. We either correlate actual shear (red) or reduced shear (green) in the calculation to highlight the small scale impact of baryons on this non linear correction. {\it Middle panel:} ratio of shear correlation functions for the two cases. {\it Bottom panel:} ratio of shear correlation functions for a raytracing that only includes rescaled DM particules or all the components.}
\label{fig:xipm}
\end{figure}

As shown in the next section, those scales remain perfectly relevant for galaxy evolution studies by means of the Galaxy-Galaxy weak lensing signal.

\section{Galaxy-Galaxy lensing}\label{sec:GGL}
Focussing further into dark matter halos, let us now investigate the yields of the simulation in terms of the galaxy-galaxy weak lensing signal. The tangential alignment of background galaxies around foreground deflectors is substantially altered by the aforementioned baryonic physics, and one also expects a strong signature in this particular lensing regime.

For a circularly symmetric mass distribution $\Sigma(R)$, one can relate shear, convergence and the mean convergence enclosed inside a radius $R$ centred on a foreground galaxy or halo as:
\begin{equation}
\bar{\kappa}(<R) = \frac{2}{R^2} \int_0^R \kappa(R') R' \der R'  = \kappa(R)+ \gamma(R)\;.
\end{equation} 
Using the definition of the critical density \eqref{eq:scrit}, one can define the excess density
\begin{eqnarray}
\Delta \Sigma(R) & = & \frac{M(<R)}{\pi R^2} - \Sigma(R) \label{eq:massexcess1} \,,\\
                 & = & \Sigma_{\rm crit} \gamma(R)\;. \label{eq:massexcess2} 
\end{eqnarray}

The previous section already showed that the lensing convergence or shear maps have adequate statistical properties, while Sect.~\ref{ssec:defl-points} showed how to use the associated deflection maps to map our lightcone galaxy catalogue into the image plane. In addition, galaxies should also get magnified when lensed. Future extensions of this work will include the  realistic photometry of the \hagn~galaxies. One can however easily account for the magnification bias by multiplying stellar masses by the magnification $\mu$, as if luminosity or flux were a direct proxy for stellar mass. In the following, we shall refer to $M_*$ for the intrinsic and $\mu M_*$ for the magnified mass proxy.

For any given source redshift, averages of the tangential shear around galaxies of any given stellar mass $M_*$ or more realistically magnified stellar mass $\mu M_*$. This is done around deflected galaxy positions.

\subsection{Comparison with CMASS galaxies}\label{ssec:GGLlow}
Let us first make a comparison of the GGL around \hagn~galaxies with the GGL excess mass profiles obtained by \citet{Lea++17} who analysed the spectroscopic CMASS sample of massive galaxies in the footprint of the CFHTLS and CS82 imaging surveys, covering $\sim 250\,{\rm deg}^2$. 
These authors paid particular attention to quantifying the stellar mass of the CMASS galaxies centred around lens redshift $z\sim0.55$. The CMASS sample is not a simple mass selection, and includes a set of colour cuts, which makes this just a broad brush comparison. These results are somewhat sensitive to the detailed distribution in stellar mass above that threshold. The sample mean mass only slightly changes with redshift but remains close to $3 \times 10^{11} \msun$.

In order to match this lens sample, we extract from the wide low redshift lightcone the galaxies in the redshift range $0.4\le z \le 0.70$, and with a stellar mass above a threshold that is chosen to match the CMASS mean stellar mass. Even though these galaxies centred around lens redshift $z\sim 0.52$ are treated as lens galaxies, they experience a modest amount of magnification (they behave like sources behind the mass distribution at yet lower redshift, see Sect.~\ref{ssec:magnif}). We thus pick galaxies satisfying $\mu M_* > 1.7 \times 10^{11} \msun$. At this stage, selecting on $M_*$ or $\mu M_*$ does not make any significant difference ($\lesssim 4\%$) because of the relatively low redshift of the lens sample. By doing so, we obtain the same sample mean stellar mass as the CMASS sample.

We now measure the mean tangential shear around those galaxies for a fiducial, unimportant, source redshift $z_{\rm s}=1$ and convert shear into excess density $\Delta \Sigma$. 
The result can be seen in Fig.~\ref{fig:ggl-prof}.  A good agreement between our predictions (OBB method, green with lighter envelope) and the observations of \citet{Lea++17} (blue dots) is found,
 further suggesting that \hagn~galaxies live in the correct massive halos ($M_{\rm h} \simeq 10^{13} \msun$), or at the very least, produce the same shear profile as CMASS galaxies around them. Note that we split the 2.25 deg field of view into 4 quadrants and used the dispersion among those areas to compute a rough estimate of model uncertainties.

On scales $R\lesssim 0.2 \hmMpc$, the shear profile is 10-15\% above the observations. Answering whether the discrepancy is due to faulty subgrid baryonic physics, a missing cosmological ingredient (or not perfectly adequate cosmological parameters) or leftover systematics in the data will certainly require more GGL observations, possibly combined with yet smaller scale strong lensing and kinematical data \citep[e.g.][]{Sonnenfeld++18}. Small scale GGL is definitely a unique tool to address those issues \citep[e.g.][]{Velliscig++17}, and asserting that the galaxy-halo connection is correctly reproduced by the simulations all the way to $z\gtrsim 1$, is arguably one of the foremost goals of galaxy formation models.

 Fig.~\ref{fig:ggl-prof} also shows our GGL results for the same population of lenses at the same redshift but as inferred from the SPL method (solid black) which allows to split the total lensing signal into its dark matter (blue and baryonic components (red). First of all, we do see a remarkable agreement between the two methods for the total lensing signal, except on scales $\gtrsim 2 \Mpc\sim 5\arcmin$ where differences start exceeding the percent level. 
 As already mentioned in the previous section, this is due to inaccuracies of the Fourier transforms performed with the SPL method. 
 We can however use this latter technic to compare the contribution of DM and baryons (stars+gas). Clearly, the total and DM profiles look very similar beyond $\sim 0.2\Mpc$ up to a $\sim17\%$ renormalisation of the matter density. It is only below those scales that cooled baryons (stars) start playing a substantial contribution. We predict an equal contribution of DM and stars to the total shear signal near a radius $\sim15\kpc$. We refer the reader to \citet{Peirani++17} for further details about the innermost density profiles around \hagn~galaxies in the context of the cusp-core problem.

\begin{figure}[h]
\includegraphics[width=0.49\textwidth]{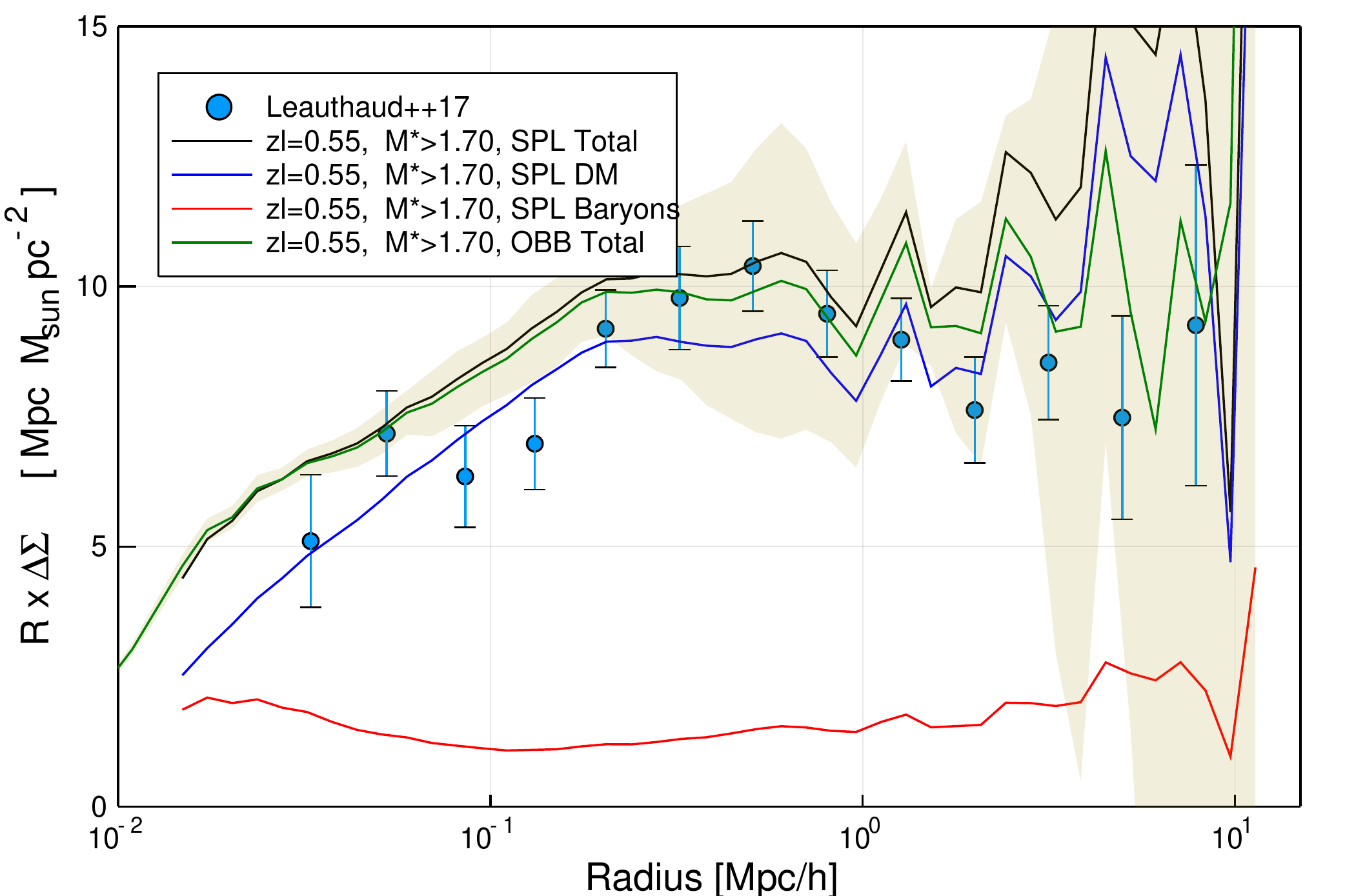}  
\caption{Comparison of the GGL tangential shear signal around $z=0.55$ \hagn ~galaxies (green curve surrounded by light-green ``ribbon'') and the GGL observations of \citet{Lea++17} (blue dots with error bars). Units are all physical (and not comoving!). Model uncertainties in the simulation past lightcone are roughly estimated by splitting the 2.25 deg wide field of view into 4 quadrants. They may be underestimated beyond $1\hmMpc$. Cuts in stellar mass are expressed in units of $10^{11} \msun$. Black, blue, and red curves show the GGL shear signal predicted with the SPL method for the total, DM, and baryonic mass distributions respectively. For clarity uncertainties are omitted. They are similar to the OBB method case (green).} 
\label{fig:ggl-prof}
\end{figure}

\subsection{High redshift magnification bias}\label{ssec:magnif}
For $z_{\rm l}\gtrsim 0.6$, the lens population starts being lensed by yet nearer structures. This can lead to a magnification bias, which was studied by \citet{Z+H08}. 

The spatial density of a lensed population of background sources can be enhanced or decreased by magnification as light rays travel through over- or under-dense sight-lines \citep[eg][]{moessner98b,moessner98a,menard02,scranton05}. Furthermore, the fraction of sources that are positively or negatively magnified depends on the slope of the luminosity function of the population. If it is very steep (typically the bright end of a population) one can observe a dramatic increase of the number of bright lensed objects. These deflectors appear brighter than they actually are. Fig.~\ref{fig:ggl-mubias} shows the mean magnification experienced by \hagn~lightcone galaxies above a given stellar mass threshold (mimicking a more realistic flux limit) as a function of redshift and minimum mass.
The upper panel does not take into account the effect of magnification bias whereas the lower panel does. 
The ones that are consistently magnified and pass a given threshold  (bottom panel) are slightly magnified on average whereas the top panel only shows a tiny constant $\mu\sim 1-3\%$ systematic residual magnification. This residual excess does not depend wether the SPL or OBB method are used, or whether we properly integrate rays or use the Born approximation. This is likely due to the replicates of the simulation box filling up the lightcone which slightly increase the probability of rays leaving an over-dense region to cross other over-dense regions on their way to the observer. This residual magnification is however tiny for sight-lines populated by galaxies and completely vanishes for rays coming for random positions.

At face value, one can see that the massive end of the galaxy stellar mass function is significantly magnification-biased. A $\sim 8\%$ effect for galaxies at $0.6\le z \le 1.2$ and $M_* \gtrsim 2 \times 10^{11} \msun$ is typical. It can be as high at $\sim 20-50\%$ at $1.5\le z \le 2$ for $\mu M_* \gtrsim 3 \times 10^{11} \msun$. A thorough investigation of the impact of this magnification bias when trying to put constraints on the high end of the $z\gtrsim 2$ luminosity function from observations is left for a forthcoming paper.

\begin{figure}[h]
\includegraphics[width=0.49\textwidth]{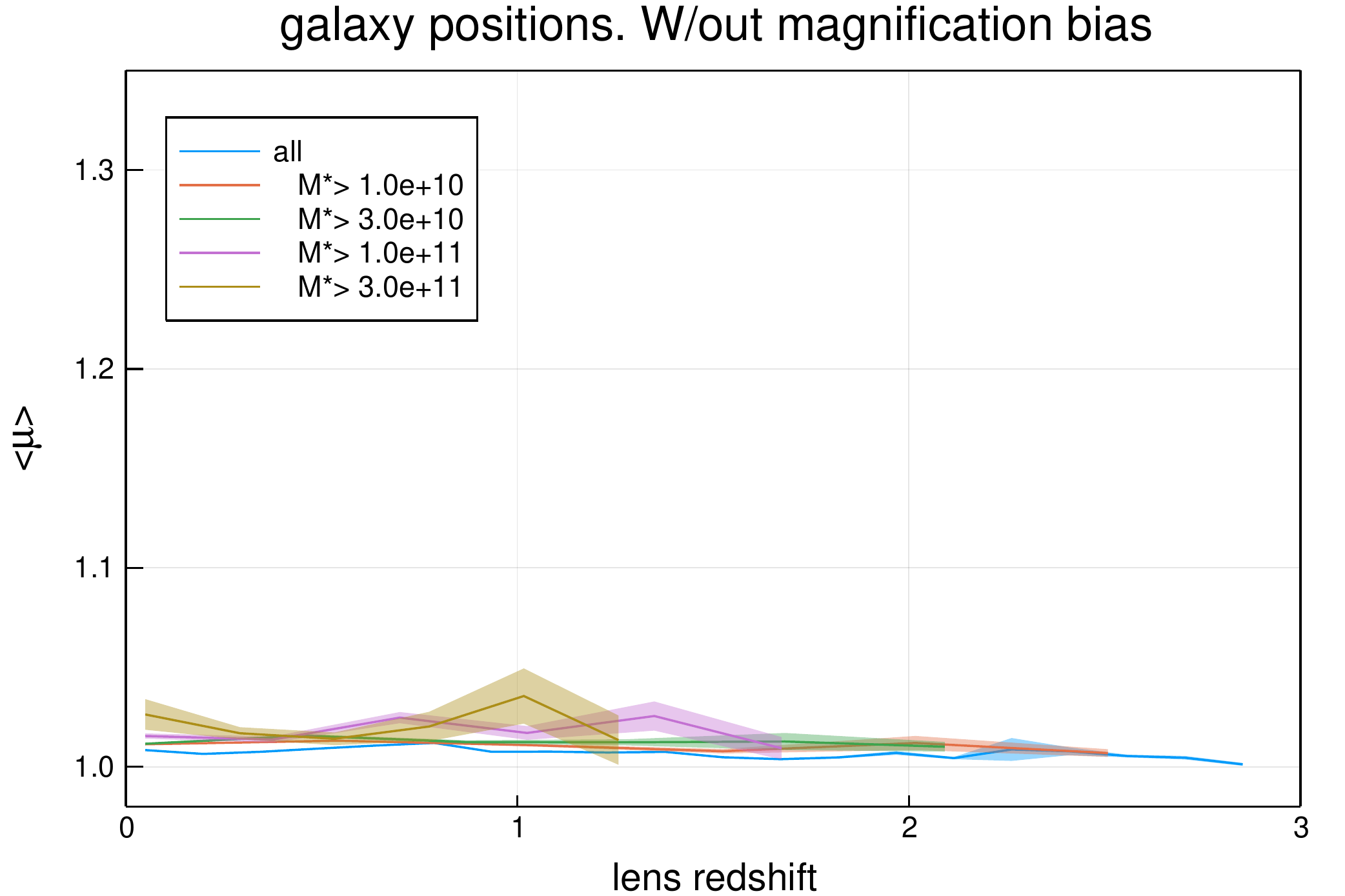}
\includegraphics[width=0.49\textwidth]{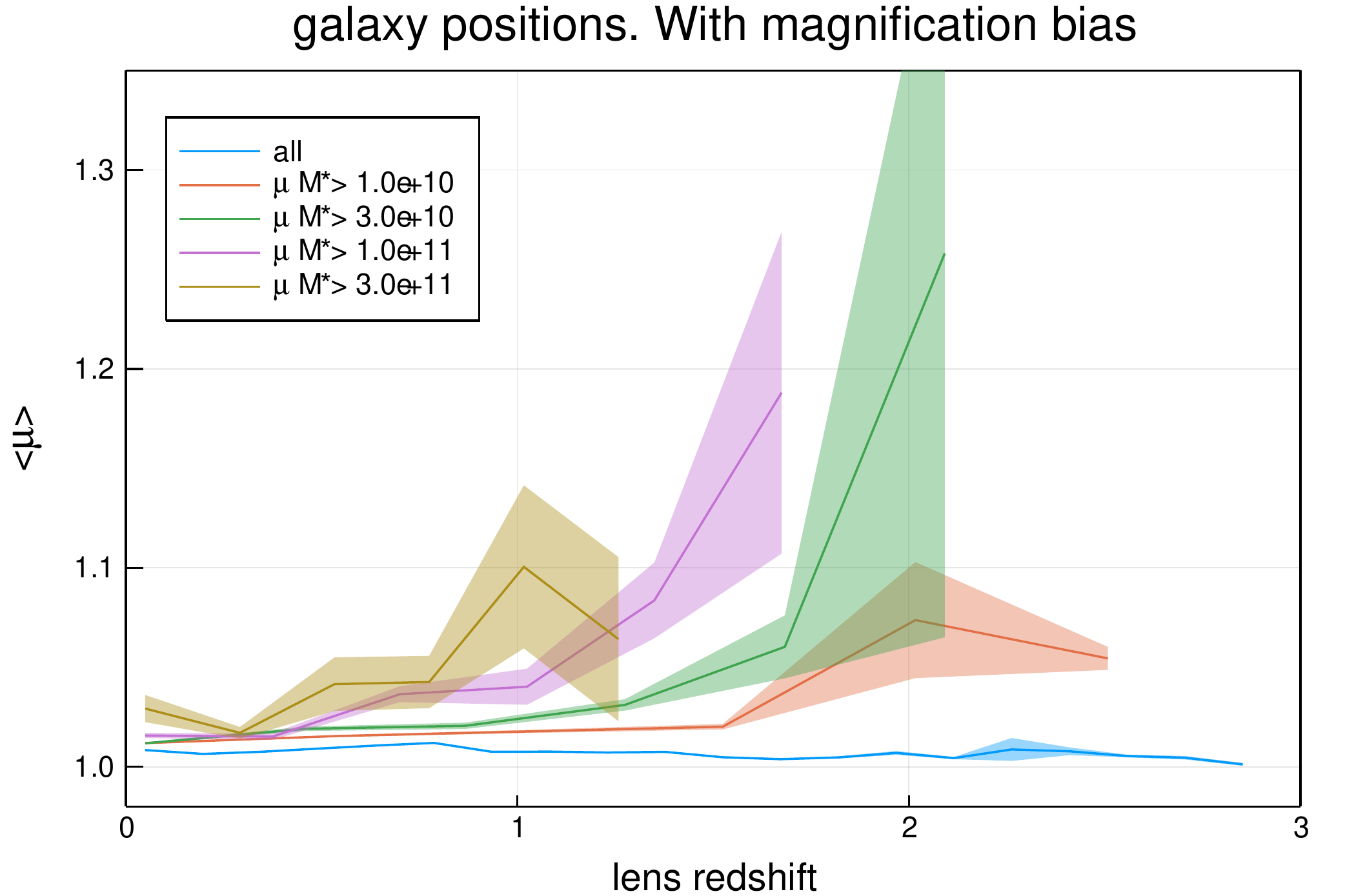}  
\caption{Average magnification experienced by presumably foreground deflectors accounting {\it (bottom)} or not {\it (top)} for magnification bias effect which mostly affects the rapidly declining high end of the stellar mass function. Without magnification bias, a flat nearly unity mean magnification at all redshifts is recovered to within $\sim1\%$. When the magnification bias is turned on, as expected in actual observations, no rapid rise is found ($\sim 10\%$ at $z\sim 1$ for the most massive/luminous galaxies). Cuts in stellar mass are expressed in units of $10^{11} \msun$.}
\label{fig:ggl-mubias}
\end{figure}

Taking magnification bias into account, let us now explore three fiducial populations of massive deflectors to highlight the changes induced on projected excess density profiles. The first population consists in the aforementioned CMASS galaxies at $z=0.54$ and $\mu M_* \ge 1.7 \times 10^{11} \msun$, the second case simply corresponds to the same  lower limit on the mass but pushed to $z=0.74$. In both cases, the excess density is measured for source redshift $z_{\rm s}=0.8$. The last lens sample corresponds to the population of H$\alpha$ emitters in the $0.9\le z\le 1.8$ redshift range that will be detected by the Euclid slit-less grism spectrograph above a line flux of $\sim 2 \times 10^{-16} \, {\rm erg\, s^{-1}\, cm^{-2}}$. One expects about 2000 such sources per square degree; therefore  the 2000 most massive \hagn~lightcone sources are picked in that redshift intervalle to crudely mimic an H$\alpha$ line flux selection. To account for magnification bias, the selection is made on $\mu M_*$, too, and the source redshift for this populations is set to $z_{\rm s}=2$. Results for these three populations can be seen in the top panel of Fig.~\ref{fig:ggl-prof-highz}, where we distinguish the excess density profiles accounting (dotted) or not (solid) for magnification. As anticipated, no significant change is obtained for the $z=0.54$ CMASS-like sample (green) but differences are more noticeable as lens redshift increases and on large scales ($R \gtrsim 1 \Mpc$), we observe a $20-50\%$ increase in $\Delta \Sigma$, consistent with the large scale linear scale-invariance bias model used by \citet{Z+H08}. Between $z=0.54$ and $z=0.74$, galaxies of the same mass seem to live in halos of the same mass (very little evolution of the $M_* - M_{\rm h}$ relation), leading to no evolution of $\Delta \Sigma$ below $\sim 200 \kpc$. The only difference occurs further out where the 2-halo term starts to be important in this galaxy-mass correlation function. There, galaxies of the same mass at $z=0.54$ and $z=0.74$ live in rarer excursions of the initial density field, and are thus more highly biased leading to an increase of $\Delta \Sigma$ on large scale. For the Euclid-like distant lens population, the trend is similar and the amplitude of the magnification bias effect would suggest a bias of the lens population about 30\% higher than it really is. 

\begin{figure}[h]
\includegraphics[width=0.49\textwidth]{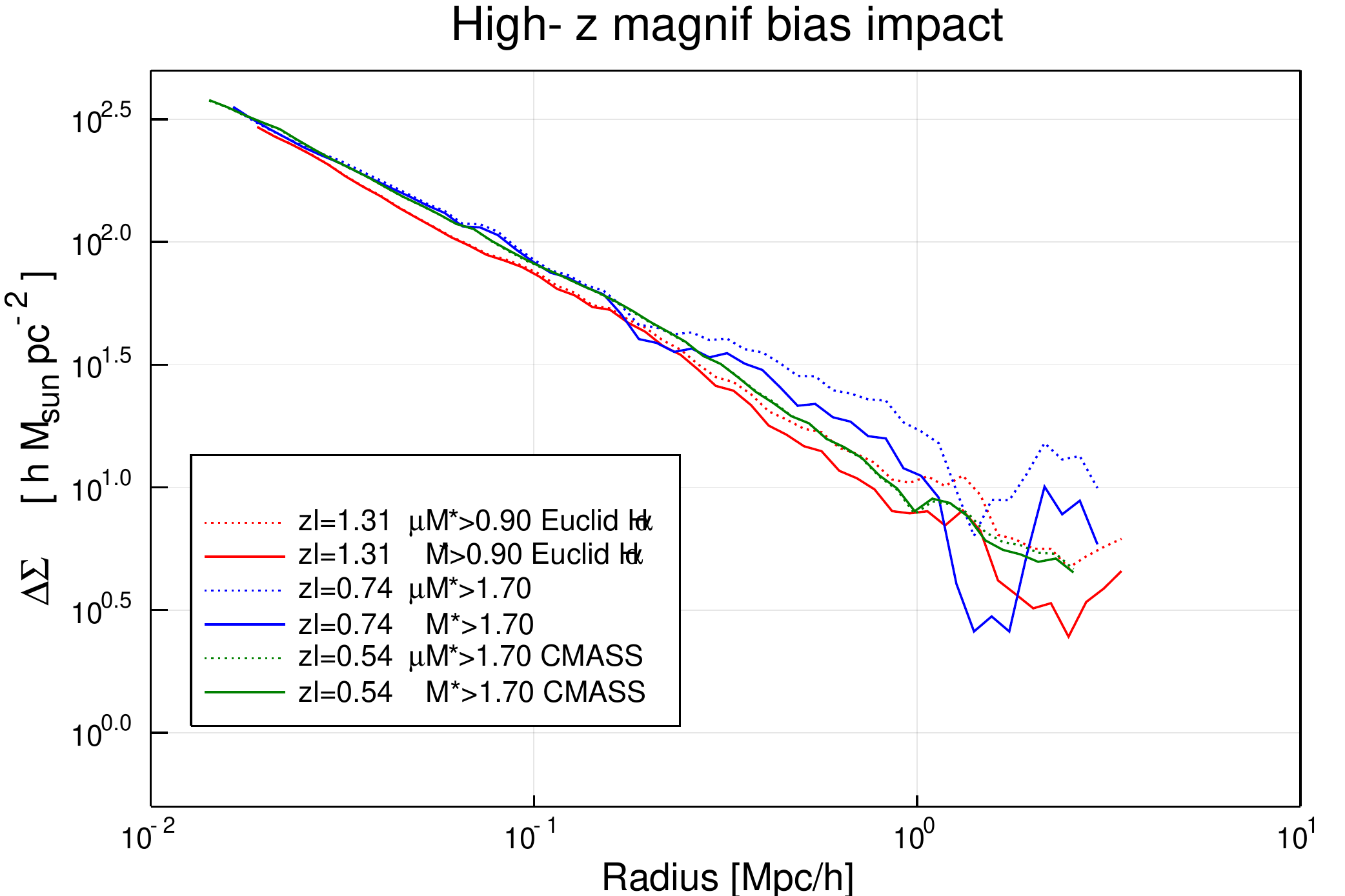}
\includegraphics[width=0.49\textwidth]{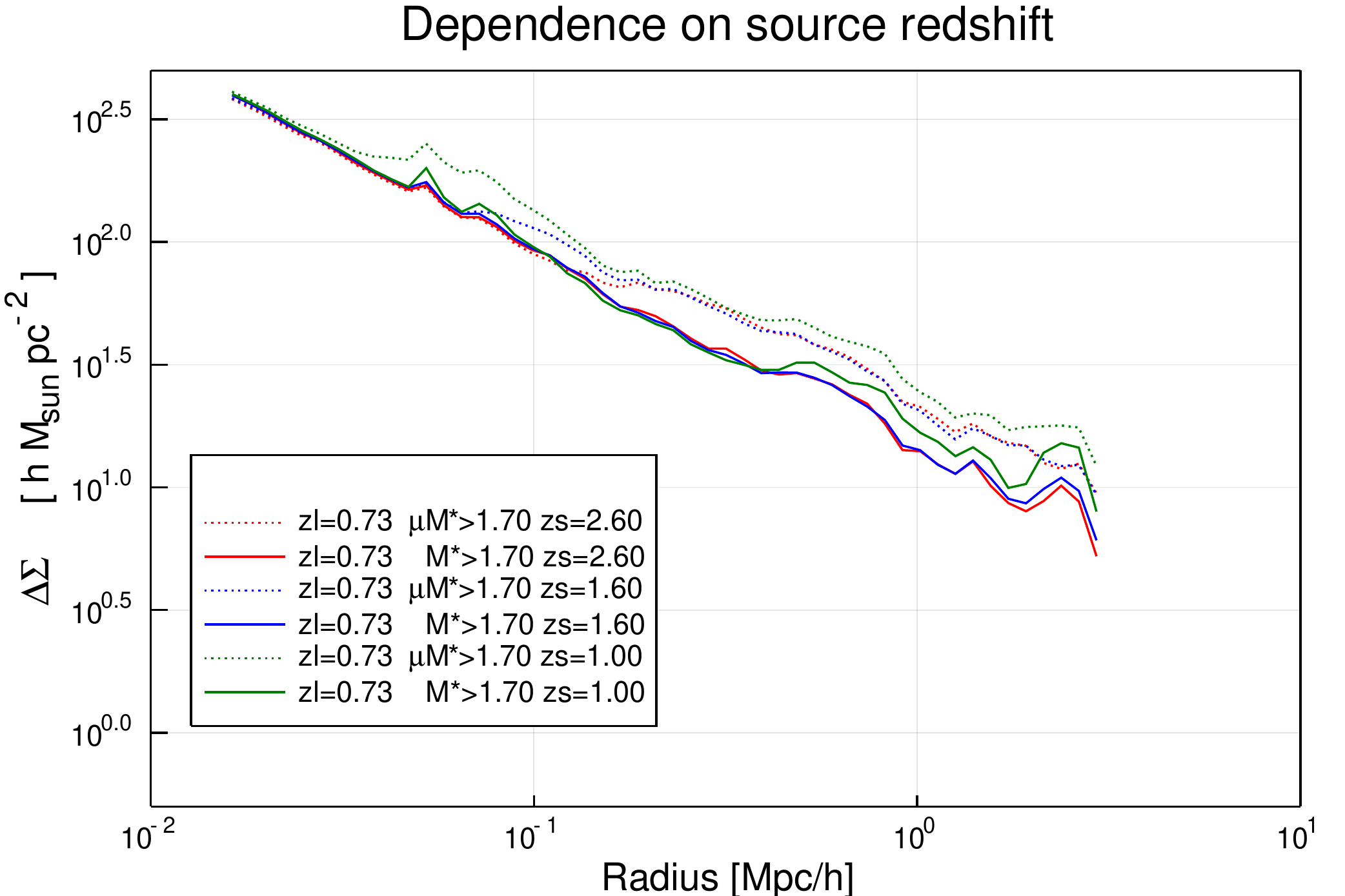}
\caption{{\it Upper panel:} Effect of magnification bias on GGL  for several high-z fiducial lens samples showing an increase of excess density $\Delta \Sigma$ (or tangential shear) for $R \gtrsim 1 \Mpc$. Solid curves ignore the magnification whereas dotted lines account for it.  {\it Lower panel:} Dependence of this effect on the source redshift. In both panels, cuts in stellar mass are expressed in units of $10^{11} \msun$.}
\label{fig:ggl-prof-highz}
\end{figure}

The lower panel of Fig.~\ref{fig:ggl-prof-highz} shows the evolution  of the magnification bias induced excess density profile  with source redshift for massive deflectors at $z=0.74$. In principle, according to equation~\eqref{eq:massexcess2}, the excess density should not depend on source redshift. However, magnification bias favours the presence of over-densities in front of deflectors. The response of distance sources carrying shear to these over-densities will depend on the source redshift in a way that is not absorbed by equation~\eqref{eq:massexcess2}. Hence,  a scale dependent distortion of the profiles is observed. The closer the source redshift from the deflector, the smaller the scale it kicks in. As already stressed by \citeauthor{Z+H08}, this hampers a direct application of shear-ratio tests with high redshift deflectors \citep[eg][]{J+T03}.


\section{Summary \& future prospects}\label{sec:conclusions}

Using two complementary methods to project the density or gravitational acceleration field from the \hagn~lightcone,  we propagated light rays and derived various gravitational lensing observables   in the simulated field of view. The simulated area is $2.25$ deg$^2$ out to $z=1$ and $1$ deg$^2$ all the way to $z=7$.  The effect of baryons on the convergence angular power spectrum $P_\kappa(\ell)$ was quantified, together with the two-point shear correlations $\xi_\pm(\theta)$ and the galaxy-galaxy lensing profile around massive simulated galaxies.

For cosmic shear, the inclusion of baryons induces a deficit of power in the convergence power spectrum of order 10\% for $10^3 < \ell < 10^4$ at $z_{\rm s}=0.5$. The amplitude of the distortion is about the same at $z_{\rm s}=1$ but is slightly shifted to roughly twice as high $\ell$ multipole values. On yet higher multipoles, the cooled baryons, essentially in the form of stars, produce a dramatic boost of power, nearly a factor $2$ for $\ell \sim 10^5$. As emphasised in \citep{Chisari++18}, it is worth stressing that detailed quantitative statements on such small angular scales may still depend on the numerical implementation of baryonic processes.

For  Galaxy-Galaxy lensing,  the projected excess density profiles for a sample of simulated galaxies consistent with the CMASS sample at $z\sim0.52$  \citep[analysed by][]{Lea++17} were  found to be in excellent agreement.
To properly analyse this signal around high redshift deflectors,   the magnification bias affecting the bright end of a population of distant galaxies was carefully taken into account, showing a large scale increase of the signal as high as 30\% beyond 1 Mpc for lenses at $z\gtrsim 1$. This kind of effect is particularly pronounced for future samples of distant deflectors, such as the spectroscopic Euclid sources detected based on their H$\alpha$ line intensity.

\citet{Peirani++18} already showed that the innermost parts of \hagn~galaxies are consistent with strong lensing observations of \citet{Sonnenfeld++13} and \citet{New++13,New++15} at $z_{\rm lens}\lesssim 0.3$. We intend to make more predictions on the optical depth for strong lensing in the \hagn~lightcone with our implemented raytracing machinery. Likewise, in a forthcoming paper we will present the results of the deflection field applied to simulated images derived from the light emitted by the stars produced in the simulation, hence enabling the possibility to measure lensing quantities (shear, magnification...) in the very same way as in observations: shape measurement in the presence of noise, Point Spread Function, pixel sampling, photometric redshift determinations, realistic galaxy biasing and more generally directly predicted galaxy-mass relation, and also the intrinsic alignment of galaxies and their surrounding halos \citep{Codis2015,2015MNRAS.454.2736C,2016MNRAS.461.2702C}.


\begin{acknowledgements}
The authors would like to thank D. Aubert for making his SPL code available to us. We acknowledge fruitful discussions with K. Benabed, S. Colombi, M. Kilbinger and S. Prunet in early phases of the project. We also thank G. Lavaux, Y. Rasera and M-A Breton for stimulating interactions around this project. We are also thankful to A. Leauthaud for constructive comments about the comparison with her GGL lensing results. CL is supported by a Beecroft Fellowship.
This work was supported by the Agence Nationale de la Recherche (ANR) as part of the SPIN(E) ANR-13-BS05-0005 \href{http://cosmicorigin.org}{http://cosmicorigin.org})  ERC grant 670193, and AMALGAM ANR-12-JS05-0006 projects, and by the Centre National des Etudes Spatiales (CNES). NEC is supported by a RAS research fellowship.
This research is also funded by the European Research Council (ERC) under the Horizon 2020 research and innovation programme grant agreement of the European Union: ERC-2015-AdG 695561 (ByoPiC, https://byopic.eu).
This work has made use of the Horizon Cluster hosted by the Institut d'Astrophysique de Paris. We thank S.~Rouberol for running  the cluster smoothly  for us.
\end{acknowledgements}

\bibliographystyle{aa}
\bibliography{hagn_lc_lens.bbl}

\end{document}